\documentclass[a4paper,10pt]{article}
\usepackage{ae}
\usepackage[ansinew]{inputenc}

\usepackage{lscape}

\usepackage{color}

\usepackage{amssymb,amsmath,amsthm,mathtools}

\def\bmath#1{\mbox{\boldmath $#1$}}

\theoremstyle{plain}

\newtheorem{thm}{Theorem}
\newtheorem{lem}{Lemma}

\newtheorem{Remark}{Remark}
\newtheorem{Assumption}{Assumption}

\title{The Einstein-Friedrich-nonlinear scalar field system and the 
stability of scalar field Cosmologies}
\author{Artur Alho$^{(1,2)}\thanks{E-mail: aalho@math.ist.utl.pt}$,
Filipe C. Mena$^{(1)}$\,\, and Juan A. Valiente Kroon$^{(3)}$\\\\
{\small $^{(1)}$Centro de Matem\'atica, Universidade do Minho, 4710-057 Braga,
Portugal}\\
{\small $^{(2)}$Centro de An\'alise Matem\'atica, Geometria e Sistemas Din\^amicos, Instituto Superior T\'ecnico, 1049-001 Lisboa,
Portugal}\\
{\small $^{(3)}$School of Mathematical Sciences, Queen Mary, University of London, Mile End, London E1 4NS, U.K.}}

\begin{document}

\maketitle

\begin{abstract}
A frame representation is used to derive a first order quasi-linear
symmetric hyperbolic system for a scalar field minimally coupled to
gravity. This procedure is inspired by similar evolution equations
introduced by Friedrich to study the Einstein-Euler system.  The
resulting evolution system is used to show that small nonlinear perturbations of expanding 
Friedman-Lema\^itre-Robertson-Walker backgrounds, with scalar field potentials satisfying 
certain future asymptotic conditions, decay exponentially to zero, in synchronous time. 
\end{abstract}

\section{Introduction}
An important problem of classical mathematical cosmology concerns the
asymptotic stability of spatially homogeneous and isotropic
spacetimes. Within this class of spacetimes, those having nonlinear
scalar field sources have been extensively used to model
early and late times cosmological scenarios. In particular, scalar field cosmologies can produce 
accelerated expansion and thus constitute possible alternatives to models with a cosmological constant \cite{Ren06}.
 
Some general results about the stability and asymptotics of scalar field cosmologies have recently been proved.
Ringstr\"{o}m \cite{Rin08,Rin09} has proved that small perturbations
of the initial data of scalar field cosmological solutions to the {\it Einstein
Field Equations} (EFE) with accelerated expansion have maximal
globally hyperbolic developments that are future causally geodesically
complete. In particular, in \cite{Rin08}, stability was shown for potentials $\mathcal{V}(\phi)$, 
satisfying $\mathcal{V}(0)>0$, $\mathcal{V}^{\prime}(0)=0$ and $\mathcal{V}^{\prime\prime}(0)>0$. 
In turn, these are potentials with a positive lower bound studied in \cite{Ren04}, for non-perturbed spatially homogeneous cosmological solutions. In fact, Rendall has shown, under mild conditions, that as $t\rightarrow+\infty$, the scalar field converges to a critical
point of the potential $\mathcal{V}(\phi_{\infty})\equiv\mathcal{V}_{\infty}>0$, $\mathcal{V}^{\prime}(\phi_{\infty})=0$ 
(with $\phi_\infty$ finite or infinite), 
and the Hubble function $H$ converges exponentially to $\sqrt{\mathcal{V}_{\infty}/3}$, 
where $\mathcal{V}_{\infty}$ is interpreted as an effective positive cosmological constant \cite{Ren04}. 
In subsequent works \cite{Ren05}, Rendall considered positive potentials for which $\mathcal{V}(\phi)\rightarrow0$, 
when $\phi\rightarrow\pm\infty$, and thus $H\rightarrow0$ as $t\rightarrow+\infty$. These are, for instance, 
solutions with exponential potentials and accelerated expansion of power-law type \cite{Hal87,BB88},
as well as other potentials which produce quasi-exponential accelerated expansion \cite{Bar90}.
Stability for power law inflation was proved by Ringstr\"{o}m in \cite{Rin09} and has also been discussed 
by Heinzle and Rendall \cite{HR07}  using Kaluza-Klein reductions and the methods of Anderson \cite{And05}. 
The latter, in turn, is inspired by Friedrich's analysis of the stability of  the de Sitter spacetime \cite{Fri86}.

\medskip
\noindent
\textbf{Main result.} A natural way to analyse the stability of spacetimes is to ask whether
small perturbations, of a given solution to the EFE, asymptotically
decay to the background solution. Most approaches to this question
have been limited to the use of linear or higher-order truncated
perturbation theory, and thus, they never take fully into account the
nonlinearity of the EFE, see
e.g. \cite{Ste90,BED92,BMT02} and 
\cite{Ren09}. This type of analysis has been hampered by the
lack of a suitable formulation of the EFE for which the theory of
systems of first order hyperbolic partial differential equations can
be applied. In this article we show how to deal with this
difficulty. Our main result shows that for an ever-expanding
FLRW-nonlinear scalar field background with spatial topology
$\mathbb{T}^{3}$, i.e. $(\mathbb{T}^{3}\times [0,\infty),\bmath{g}_{\text{FLRW}},\phi)$,
and scalar field potentials $\mathcal{ V}$ satisfying the future asymptotic
conditions 
\[
\mathcal{ V}_{\infty}>0, \qquad
-\left(\frac{\mathcal{ V}^{\prime}}{d \phi/dt}\right)_{\infty}>\sqrt{\frac{\mathcal{ V}_{\infty}}{3}},
\qquad \mathcal{ V}^{\prime\prime}_{\infty}>0,
\]
 nonlinear perturbations exist and exponentially decay 
to zero, asymptotically, in synchronous time $t$. Our result is proved in an equivalent norm to the Sobolev norm
$H^{k}\left(\mathbb{T}^{3}\right)$ for $k\geq5$. 

\medskip
\noindent
\textbf{Strategy of the analysis.} In \cite{Fri96}, Friedrich has introduced a frame
representation of the vacuum EFE, see also \cite{EU99} for a similar construction. 
The evolution equations implied by this alternative representation of the equations of General
Relativity constitute a \emph{first-order quasi-linear symmetric
hyperbolic  system (FOSH)}. In general, these systems are of the form
\begin{equation}
 \mathbf{A}^{0}(\mathbf{u})\partial_{t}\mathbf{u}-\mathbf{A}^{j}(\mathbf{u})\partial_{j}\mathbf{u}=\mathbf{B}(\mathbf{u})\mathbf{u}
\label{FOSHsystem}
\end{equation}
where $\mathbf{u}=\mathbf{u}(\bmath{x},t)$ is a smooth vector-valued
function of dimension $s$ with domain in $\Sigma\times[0,T]$ where 
$\Sigma$ is a spacelike 3-dimensional manifold. Moreover, $\mathbf{A}^{0}$,
$\mathbf{A}^{j}$, $j=1,\,2,\,3$, and $\mathbf{B}$ denote smooth
$s\times s$ matrix valued-functions, such that  $\mathbf{A}^0$ and $\mathbf{A}^{j}$ are
symmetric and $\mathbf{A}^{0}$ positive definite.  The
operators $\partial_{t}$ and $\partial_{j}$ stand, respectively, for
the partial derivatives with respect to the coordinates $t\in[0,T]$ and
$(x^{j})\in\Sigma$.

The construction for vacuum spacetimes given in
\cite{Fri96}, has been extended in
\cite{Fri98,FR00} to the case of a
\emph{perfect fluid} using a Lagrangian description of the fluid flow (see also \cite{CBY01,CBY02}). In both
the vacuum and the perfect fluid cases, the introduction of a frame
formalism gives rise to extra gauge freedom. 
This freedom is associated to the evolution of the spatial frame coefficients along
the flow of the time-like frame. If one fixes conveniently this gauge
(using, for example, the Fermi gauge), one obtains a hyperbolic reduction for
the evolution equations. As a consequence, given smooth initial data
satisfying the constraints, local existence  in time and uniqueness of a
solution to the EFE can be established (see
e.g. \cite{FR00,Reu98} and also
\cite{CBY01,CBY02} for details).

A natural way of performing a stability analysis is to consider a sequence of smooth initial data sets 
$\mathbf{u}^{\epsilon}_{0}$ for the EFE satisfying the constraints
equations on a Cauchy hypersurface $\Sigma$. The sequence is assumed
to depend continuously on the parameter $\epsilon$ in such a way
that the limit $\epsilon\rightarrow0$ renders the data of the reference
solution $\mathring{\mathbf{u}}_{0}$. In particular, one can write the full
solution to the EFE as the ansatz
\begin{equation}
\label{Ansatz}
 \mathbf{u}^{\epsilon}=\mathring{\mathbf{u}}+\epsilon\breve{\mathbf{u}},
\end{equation}
where $\breve{\mathbf{u}}$ is a (nonlinear)
perturbation whose size is controlled by the parameter
$\epsilon$. Using the ansatz in equation
\eqref{FOSHsystem}, and writing
\begin{equation}
\begin{aligned}
 \mathbf{B}(\mathring{\mathbf{u}}+\epsilon\breve{\mathbf{u}})&\equiv\mathbf{B}(\mathring{\mathbf{u}})+\epsilon\mathbf{B}(\mathring{\mathbf{u}},\breve{\mathbf{u}},\epsilon),\\
\mathbf{A}^{\mu}(\mathring{\mathbf{u}}+\epsilon\breve{\mathbf{u}})&\equiv\mathbf{A}^{\mu}(\mathring{\mathbf{u}})+\epsilon\mathbf{A}^{\mu}(\mathring{\mathbf{u}},\breve{\mathbf{u}},\epsilon),\quad \mu=0,1,2,3
\end{aligned}
\label{ansatz}
\end{equation}
we are led to consider an initial value problem for the
nonlinear perturbations of the form:
\begin{eqnarray}
&&
\left(\mathring{\mathbf{A}}^{0}+\epsilon
  \breve{\mathbf{A}}^{0}\right)\partial_{t}\breve{\mathbf{u}}-\left(\mathring{\mathbf{A}}^{j}+\epsilon
  \breve{\mathbf{A}}^{j}\right)\partial_{j}\breve{{\bf
    u}}=\left(\mathring{{\bf B}}+\epsilon \breve{{\bf B}}\right)\breve{{\bf u}},\nonumber \\
&&\breve{{\bf u}}(\bmath{x},0)=\breve{{\bf u}}_{0}(\bmath{x}). 
\label{nonlinearequation} 
\end{eqnarray}
Here, the coefficients $\mathring{\mathbf{B}}\equiv\mathbf{B}(\mathring{\mathbf{u}})$ and
$\mathring{\mathbf{A}}^{\mu}=\mathbf{A}^{\mu}(\mathring{\mathbf{u}})$ in the splitting \eqref{ansatz} are defined
uniquely by the condition $\epsilon=0$. Also
\[
\breve{\mathbf{A}}^{\mu}\equiv\mathbf{A}^{\mu}(\mathring{\mathbf{u}},\breve{\mathbf{u}},\epsilon)
\]
and
\[
\breve{\bf B}\breve{\bf u} \equiv\mathbf{B}(\mathring{\mathbf{u}},\breve{\mathbf{u}},\epsilon)\mathring{\bf u} +
\mathbf{B}(\mathring{\mathbf{u}},\breve{\mathbf{u}},\epsilon)\breve{\bf u} + \breve{\bf A}^j\partial_j
\mathring{\bf u} - \breve{\bf A}^0\partial_t \mathring{\bf u},
\]
where it has been assumed that
\[
\mathring{\mathbf{A}}^{0}\partial_{t}\mathring{\mathbf{u}}-\mathring{\mathbf{A}}^{j}\partial_{j}\mathring{{\bf
    u}} =\mathring{{\bf B}}\mathring{{\bf u}}.
\]
A particular approach to the existence and stability of solutions to
the Cauchy problem \eqref{nonlinearequation}, for the case where the
coefficients of the linearized system ($\epsilon=0$) are constant
matrices, has been discussed in
\cite{KREISSBOOK,KS97,KL98,KOR98,Ort01}.  In this approach, the
asymptotic future stability of solutions follows from the existence of
eigenvalues for the non-principal part of the linearised system having
a negative real part ({\it strictly dissipative systems}).  In the
case where the system is only strongly hyperbolic, the inner product
in $L^{2}$ has to be replaced by the so-called $\mathcal{H}$-inner
product ---see \cite{KOR98}. A procedure to analyse stability in the
case of systems where $\mathring{{\bf B}}$ has vanishing eigenvalues
({\it dissipative systems}) has been given in \cite{KOR98} ---see also
\cite{KNOR97,Ort01}.

In this paper,  
we will show how these methods can be generalized to systems of
the type considered here, where the matrices $\mathring{{\bf B}}$,
$\mathring{\mathbf{A}}^{\mu}$ are not constant but depend smoothly on
time. A similar
analysis has been adopted by Reula in \cite{Reu99}, using the
Einstein-perfect fluid system of \cite{Fri98} with a positive
cosmological constant $\Lambda>0$, to prove the exponential decay of
nonlinear perturbations for a wide class of homentropic fluids in flat
{\it Friedman-Lema\^itre-Robertson-Walker} (FLRW)
backgrounds\footnote{The presence of a cosmological constant is
crucial for global existence and exponential decay, since the minimum
of the Hubble function must be strictly positive, namely
$H_{min}=\sqrt{\Lambda/3}$.}.  An advantage of this approach is that
it avoids the problem of gauge-dependence in perturbation theory and, therefore,
gauge-invariant conclusions, such as geodesic completeness, can be
inferred.

We analyse the nonlinear stability
of FLRW spacetimes with a nonlinear scalar field. To this end, we
first construct a first order symmetric hyperbolic system for the EFE
with a scalar field as the matter source. This construction is
performed by splitting the wave equation for the scalar field into two
first order equations. In our analysis, the scalar field is used 
to construct an adapted orthogonal frame, for which the energy-momentum tensor is
diagonal, independently of further gauge choices.  
Similar splittings are often used 
in the analysis of linear perturbations 
\cite{Mad88,BED92,AM13,UW12}.

\medskip
\noindent
\textbf{Structure of the article.} The article is organized as follows: in Section 2, we recall Friedrich's
frame formulation of the EFE. In Section 3, we discuss some relevant
properties of scalar fields satisfying a nonlinear wave 
equation. In Section 4, we discuss the conditions under which the Einstein-Friedrich-nonlinear
scalar field system is well-posed ---in the sense that it forms a
symmetric hyperbolic system, see  Theorem \ref{Theorem:wellposedness}. 
Finally, in Section 5, we give the conditions for which there is an asymptotic exponential decay of 
small nonlinear perturbations on a FLRW-nonlinear scalar field background. This is the main result
of the paper and we summarise it in Theorem \ref{Theorem:Main}.  We use units
such that $8\pi G=c=1$.

\section{Friedrich's frame formulation of the Einstein Field Equations}
In this section, we provide a brief introduction to Friedrich's frame formulation of the Einstein field equations. The basic equation of Friedrich's construction is the contracted Bianchi identity. From the latter, it is possible to deduce hyperbolic propagation equations for the conformal Weyl tensor for a wide class of gauge choices.

\subsection{Basic definitions and notation}

In order to implement the frame formulation of the Einstein field
equations, one defines locally an orthonormal moving frame or
$\it{tetrad}$ with respect to the metric $\bmath{g}$ in an open
neighbourhood $\mathcal{U} \subset \mathcal{M}$.  The frame is a set
$\left\{\bmath{e}_{a}\right\}$ of linearly independent vector fields
in the tangent space $T_{p}\left(\mathcal{M}\right)$ at each point
$p\in\mathcal{U}$ such that
\begin{equation}\label{orthon}
 \bmath{g}\left(\bmath{e}_{a},\bmath{e}_{b}\right)=\eta_{ab}, \quad a,
 \, b=0,1,2, 3,
\end{equation}
where $\eta_{ab}=\text{diag}\left(-1,1,1,1\right)$ and latin letters (except for the $i,j$) are used for frame indices. The {\it norm} of a vector field, $\bmath{v}\in T_{p}\left(\mathcal{M}\right)$, in an orthonormal frame is defined as
\begin{equation*}
 |\bmath{v}|^{2}\equiv\bmath{g}\left(\bmath{v},\bmath{v}\right)=v^{a}v^{b}\eta_{ab}
\end{equation*}
and, in terms of a coordinate basis set $\left\{\partial_{\alpha}\right\}$, we have $\bmath{e}_{a}=e_{a}^{\,\,\mu}\partial_{\mu}$. Condition \eqref{orthon} gives
\begin{equation*}
 \eta_{ab}=e_{a\,}^{\,\,\mu}e_{b}^{\,\,\nu}g_{\mu\nu},
\end{equation*}
wherein $\mu,\nu=0,1,2,3$. The \emph{frame commutator} is written as
\begin{equation}\label{commutatorframes}
 [\bmath{e}_{a},\bmath{e}_{b}]=c^{c}_{\,\,ab}\bmath{e}_{c},
\end{equation}
where $c^{c}_{\,\,ab}$ are the \emph{structure coefficients}.  The \emph{dual basis} or \emph{coframe} is the set of linear forms
$\{\bmath{\theta}^{b}\}$ belonging to the dual space $T^{*}_{p}\left(\mathcal{M}\right)$ at each point $p\in\mathcal{U}$ defined by the pairing $\langle\bmath{\theta}^{b},\bmath{e}_{a}\rangle=\delta_{a}{}^{b}$. In terms of the dual basis, we can write condition \eqref{orthon} as
\begin{equation*}
 \bmath{g}=-\left(\bmath{\theta}^{0}\right)^{2}+\sum^{3}_{a=1}\left(\bmath{\theta}^{a}\right)^{2}.
\end{equation*}
The \emph{spacetime (Levi-Civita) connection}, in an orthonormal basis, is defined by
\begin{equation*}
\bmath{\nabla}_{a}\bmath{e}_{b}\equiv\gamma^{c}_{\,\,\,ba}\bmath{e}_{c},
\end{equation*}
where $\gamma^{c}_{\,\,ba}$ are the \emph{connections coefficients} and the covariant derivative of a tensor, in $\mathcal{M}$, can be written as
\begin{eqnarray*}
&& \bmath{\nabla}_{a}v_{q_{1}...q_{s}}{}^{p_{1}...p_{r}}=\bmath{e}_{a}\left(v_{q_{1}\cdots q_{s}}{}^{p_{1} \cdots p_{r}}\right)+\gamma_{\,\,\,fa}^{p_{1}}v_{q_{1} \cdots q_{s}}{}^{f \cdots p_{r}}+ \cdots \\
&& \hspace{4cm} \cdots+\gamma_{\,\,\,fa}^{p_{r}}v_{q_{1} \cdots q_{s}}{}^{p_{1} \cdots f}-\gamma_{\,\,\,q_{1}a}^{f}v_{f \cdots q_{s}}{}^{p_{1} \cdots p_{r}}-\cdots-\gamma_{\,\,\,q_{s}a}^{f}v_{q_{1} \cdots f}{}^{p_{1} \cdots p_{r}}.
\end{eqnarray*}
The torsion free and metric compatibility conditions imply, respectively, that
\begin{equation*}
 c^{c}_{\,\,ab}=\gamma^{c}_{\,\,ba}-\gamma^{c}_{\,\,ab}, \quad \gamma^{e}_{\,\,ba}\eta_{ec}+\gamma^{e}_{\,\,ca}\eta_{eb}=0,
\end{equation*}
while the equations for the frame coefficients
$\left\{e_{a}^{\,\,\mu}\right\}$ are given by equation
\eqref{commutatorframes} in terms of the connection coefficients. In
turn, equations for the connection coefficients are obtained from the
Ricci identity
\begin{equation}
 R^{a}_{\,\,bcd}=\bmath{e}_{c}\left(\gamma^{a}_{\,\,bd}\right)-\bmath{e}_{d}\left(\gamma^{a}_{\,\,bc}\right)+\gamma^{a}_{\,\,fc}\gamma^{f}_{\,\,bd}-\gamma^{a}_{\,\,fd}\gamma^{f}_ {\,\,bc}-\gamma^{a}_{\,\,bf}\left(\gamma^{f}_{\,\,dc}-\gamma^{f}_{\,\,cd}\right).
\label{RicciIdentity}
\end{equation}
The Riemann tensor can be decomposed in terms of the conformal Weyl
tensor $\bmath{C}$ and the Schouten tensor $\bmath{S}$ as
\begin{equation}\label{decompositionRiemann}
 R^{a}{}_{bcd}=C^{a}{}_{bcd}+\delta^{a}{}_{[c}S_{d]b}-\eta_{b[c}S_{d]}{}^a.
\end{equation}
For future use, we introduce the \emph{Friedrich tensor} $\bmath{F}$ via
\begin{equation}\label{F}
 F_{abcd}\equiv {C}_{abcd}-\eta_{a[c}S_{d]b},
\end{equation}
and its dual with respect to the last pair of indices
\begin{equation}\label{Fdual}
 {^{\star}}F_{abcd}={^{\star}}{C}_{abcd}+\frac{1}{2}S_{pb}\epsilon^{p}{}_{acd},
\end{equation}
where $\epsilon_{abcd}$ is the usual Levi-Civita totally antisymmetric
symbol with $\epsilon_{0123}=1$. In terms of the Friedrich tensor, one
finds that the contracted Bianchi identities read
\begin{equation}\label{BianchiIdentities}
 \nabla_{a}F^{a}_{\,\,bcd}=0,\quad\nabla_{a}{^{\star}}F^{a}_{\,\,bcd}=0.
\end{equation}
%
%
\subsection{Orthonormal decomposition of the field equations}
The equations of Friedrich's frame formulation of the Einstein field equations are given by \eqref{commutatorframes},
\eqref{RicciIdentity} and \eqref{BianchiIdentities}, together with the decomposition \eqref{decompositionRiemann}. The independent variables of the system are therefore
\begin{equation*}
 \left(e_{a}{}^\mu,\gamma^{a}{}_{bc},C^{a}{}_{bcd},S_{bc}\right).
\end{equation*}
In what follows, we shall decompose the equations and relevant tensors in terms of their parallel and orthogonal components with respect to the time-like frame. We write
$\bmath{N}\equiv \bmath{e}_{0}$ and set 
\begin{equation*}
\bmath{N}=N^{a}\bmath{e}_{a},\quad N^{a}=\delta_0{}^a,
\end{equation*}                                                 
where $N_{a}=-\delta_a{}^0$, in our signature. In terms of these objects, tensor fields which are orthogonal to the timelike frame-vector are defined by
\begin{equation*}
 T_{a_{1}\dots a_{p}\dots a_{q}}N^{a_{p}}=0,\quad p=1, 2,\ldots, q.
\end{equation*}
Defining the projector onto the orthogonal 3-subspaces
\begin{equation*}
 h_{ab} \equiv \eta_{ab}+N_{a}N_{b},
\end{equation*}
where $h_{a}{}^{c}=\eta^{bc}h_{ab}$, the \emph{spatial covariant derivative} is then given by 
\begin{equation*}
 D_{a}T_{q_{1}\cdots q_{r}}=h_{a}{}^{b}h_{q_{1}}{}^{p_{1}}\cdots h_{q_{r}}{}^{p_{r}}\nabla_{b}T_{p_{1}\cdots p_{r}}.
\end{equation*}
In particular, one has 
\begin{equation*}
 D_{a}h_{bd}=0,\quad D_{a}\epsilon_{bcd}=0,
\end{equation*}
where $\epsilon_{bcd}$ is the spatial Levi-Civita symbol and the indices run from $1$ to $3$. In order to further proceed with the geometric decomposition one defines the \emph{acceleration vector} by
\begin{equation*}
 \bmath{a}\equiv \bmath{\nabla}_{0}\bmath{e}_{0}=\gamma^{p}{}_{00}\bmath{e}_{p}, \qquad p=1,2,3.
\end{equation*}
It follows then that $a^{p}=\gamma^{p}_{\,\,\,00}$ or equivalently, $a_{p}=\gamma^{0}_{\,\,\,p0}$. We will also consider the so-called
\emph{Weingarten map} given by 
\begin{equation*} 
 \bmath{\chi}(\bmath{e}_{a})\equiv \bmath{\nabla}_{a}\bmath{e}_{0}=\gamma^{p}_{\,\,\,0a}\bmath{e}_{p},\quad a,p=1,2,3,
\end{equation*}
with $\chi_{a}{}^{p}=\gamma^{p}{}_{0a}$.  The tensor $\chi_{ab}$ can be written in terms of its irreducible parts as
\begin{equation*}
 \chi_{ab}=\gamma^{0}{}_{ba}=\left(\chi^{ST}\right)_{ab}+\frac{1}{3}\chi h_{ab}+\left(\chi^{A}\right)_{ab},
\end{equation*}
where $(\chi^{ST})_{ab}$, $\chi$, $(\chi^{A})_{ab}$ denote, respectively, its symmetric trace-free, trace and antisymmetric
parts. If the flow of $\bmath{e}_{0}$ is hypersurface orthogonal, then one has that $(\chi^{A})_{ab}=0$ and that
\begin{equation}
 \frac{1}{2}\pounds_{N}h_{ab}=\chi_{(ab)}=(\chi^{ST})_{ab}+\frac{1}{3}\chi h_{ab},
\label{LieDer3metric}
\end{equation}
where $\pounds_{N}$ denotes the Lie derivative along $\bmath{N}$ and $\nabla_{a}N^{p}=-N_{a}a^{p}+\chi_{a}{}^{p}$. Finally, the 4-dimensional Levi-Civita symbol is also decomposed using 
\begin{equation*}
 \epsilon_{abcd}=2\epsilon_{ab[c}N_{d]}-2N_{[a}\epsilon_{b]cd}.
\end{equation*}
Now, defining $\tilde{F}_{bcd}\equiv \nabla_{a}F^{a}{}_{bcd}$, it follows that the first contracted Bianchi identity can be written as
\begin{equation}
 \tilde{F}_{bcd}=N_{b}\left[\bar{\tilde{F}}_{0c0}N_{d}-\bar{\tilde{F}}_{0d0}N_{c}\right]+2\bar{\tilde{F}}_{b0[c}N_{d]}-N_{b}\bar{\tilde{F}}_{0cd}+\bar{\tilde{F}}_{bcd}=0,
\label{GeomDecBI}
\end{equation}
where contractions with $\bmath{N}$ are denoted by the index $0$, and the bar $\;\bar{}\;$ indicates that the remaining indices are
spatial. For example, $\bar{\tilde{F}}{}_{b0d}\equiv h_{b}{}^{q}N^{r}h_{d}{}^{s}\tilde{F}_{qrs}$. Given the vector $\bmath{N}$, the Weyl tensor is uniquely determined through its \emph{electric} and \emph{magnetic} parts defined, respectively, by
\begin{equation*}
 E_{ab}\equiv h_{a}{}^{q}h_{b}{}^{d}N^{p}N^{c}C_{pqcd},\quad B_{bd}\equiv h_{b}{}^{p}h_{d}{}^{q}N^{a}N^{c} {^{\star}}C_{apcq}.
\end{equation*}
In terms of the latter, the Weyl tensor and its dual can be written as 
\begin{eqnarray}
&& C_{abcd}=2\left[l_{a[c}E_{d]b}-l_{b[c}E_{d]a}\right]-2\left[N_{[c}B_{d]p}\epsilon^{p}{}_{ab}+N_{[a}B_{b]p}\epsilon^{p}{}_{cd}\right]
\label{DecompositionWeyl}\\
&& {^{\star}}C_{abcd}=2N_{[a}E_{b]p}\epsilon^{p}{}_{cd}-4E_{p[a}\epsilon_{b]}{}^{p}{}_{[c}N_{d]}-4N_{[a}B_{b][c}N_{d]}-B_{pq}\epsilon^{p}{}_{ab}\epsilon^{q}{}_{cd},
\label{DecompositionWeylDual}
\end{eqnarray}
where $l_{ab}\equiv h_{ab}+N_{a}N_{b}$.

\section{Nonlinear scalar fields in the frame formalism}
In this section, we introduce a description of nonlinear scalar fields which is particularly well adapted to 
the present analysis.
\subsection{Basic equations}
In general, the energy-momentum tensor for a smooth nonlinear scalar field 
has the form 
\begin{equation*}
\bmath{T}=\bmath{\psi} \otimes \bmath{\psi}-\left(\frac{1}{2}
  |\bmath{\psi}|^{2} +\mathcal{V}(\phi)\right)\mathbf{g},
\end{equation*}
where we have defined the 1-form
\begin{equation*}
{\bmath{\psi}}\equiv \mathbf{d}\phi.
\end{equation*}
Accordingly, we define
\begin{equation}\label{definitionpsi}
 \psi_{a}\equiv {\bmath{\psi}}\left(\bmath{e}_{a}\right)=(\psi,\bar{\psi}_{a}),
\end{equation}
where we have written 
\begin{equation}
 \psi\equiv \psi_{0}=\pounds_{N}\phi
\label{psi_{0}_frame}
\end{equation}
and
\begin{equation}
\bar{\psi}_{a}\equiv h_{a}{}^b \psi_{b}=D_{a}\phi.
\end{equation}
The components of the energy-momentum tensor $\bmath{T}$, with respect to the tetrad $\{\bmath{e}_a\}$, are then given by
\begin{equation}
\label{stress}
 T_{ab}=\psi_{a}\psi_{b}-\left(\frac{1}{2} |\bmath{\psi} |^{2}+\mathcal{V}(\phi)\right)\eta_{ab},
\end{equation}
while its trace is
\[
T=-|\bmath{\psi} |^{2}-4\mathcal{V}(\phi).
\]
The Einstein field equations,
imply for the components of the Ricci tensor, that
\begin{equation*}
 R_{ab}=\psi_{a}\psi_{b}+\mathcal{V}(\phi)\eta_{ab},
\end{equation*}
while the Ricci scalar is given by
\[
R=-T=|\bmath{\psi}|^{2}+4\mathcal{V}(\phi).
\]
From these expressions, it follows that the components of the Schouten tensor with respect to the frame $\{\bmath{e}_a\}$ are given by
\begin{equation*}
 S_{ab}=\psi_{a}\psi_{b}-\frac{1}{3}\left(\frac{1}{2}|\bmath{\psi}|^{2}-\mathcal{V}(\phi)\right)\eta_{ab}.
\end{equation*}

\subsection{Gauge considerations}
In order to construct an adapted frame to our particular problem, 
we let $\bmath{\psi}\equiv\alpha\bmath{e}_{0}$. It follows that
\begin{equation}
 \quad\psi^{a}=\alpha\delta^{a}_{0},
\label{fixingpsi}
\end{equation}
so that
\[
\alpha=-\psi\quad\text{and}\quad D^{a}\phi=0.
\]
Accordingly, 
\begin{equation}
 |\bmath{\psi}|^{2}=\mathbf{g}\left(\bmath{\psi},\bmath{\psi}\right)=\alpha^{2}\eta_{00}=-\alpha^{2},
 \quad\alpha=\pm\sqrt{-|\bmath{\psi}|^{2}}.
\label{definitionalpha}
\end{equation}
If the vector $\bmath{\psi}$ is taken to be future oriented, then one
must choose $\alpha$ to be positive. In terms of a coordinate basis,
the latter implies
\begin{equation}
 \psi^{\mu}=\alpha e_{0}{}^{\mu}=-\psi e_{0}{}^{\mu}, \quad  e_{0}{}^{\mu}=\frac{\nabla^{\mu}\phi}{\sqrt{-|\bmath{\psi}|^{2}}}
\label{fixingpsicoordinates}
\end{equation}
and
\begin{equation}
D_{a}\phi=0, \quad  \bar{e}_{a}{}^\mu\nabla_{\mu}\phi=0 .
\label{constraint_Da_phi}
\end{equation}
With this choice, we have
\[ 
\psi_{a}=-\psi N_{a},
\]
 and therefore
\begin{eqnarray}
 &&T_{ab}=\left(\frac{1}{2}\psi^{2}+\mathcal{V}(\phi)\right)N_{a}N_{b}+\left(\frac{1}{2}\psi^{2}-\mathcal{V}(\phi)\right)h_{ab},\\
 &&S_{ab}=\frac{1}{3}\left(\frac{5}{2}\psi^{2}-\mathcal{V}(\phi)\right)N_{a}N_{b}+\frac{1}{3}\left(\frac{1}{2}\psi^{2}+\mathcal{V}(\phi)\right)h_{ab}.
\label{Shoutenscalarfield}
\end{eqnarray}
\begin{Remark}
By fixing 
 $\bmath{\psi}=\alpha\bmath{e}_{0}$, we assume that $\bmath{\psi}$ is timelike. If this is not the case, 
 then our gauge breaks and the evolution stops. We are thus considering a subset of solutions to 
 the EFEs for which this choice is valid. We note that this is a common choice in cosmology, see e.g. \cite{BED92}. 
\end{Remark}
Using equations \eqref{definitionpsi} and \eqref{definitionalpha}, the expression
for the conservation of the energy-momentum tensor takes the form
\begin{equation}\label{div_EMtensor}
\begin{aligned}
 \nabla^{a}T_{ab}=&\nabla^{a}\left(\psi^{2}N_{a}N_{b}+\left(\frac{1}{2}\psi^{2}-\mathcal{V}(\phi)\right)\eta_{ab}\right) \\
                 =&2\psi N_{b}N^{a}\left(\nabla_{a}\psi\right)+\psi^{2}\left(N_{b}\left(\nabla_{a}N^{a}\right)+N^{a}\left(\nabla_{a}N_{b}\right)\right)+\nabla_{b}\left(\frac{1}{2}\psi^{2}-\mathcal{V}(\phi)\right) \\
                =&\left(2\psi\pounds_{\text{\tiny{\bmath{N}}}}\psi+\psi^{2}\chi+\psi\frac{d\mathcal{V}}{d\phi}\right)N_{b}+\psi^{2}a_{b}+\psi\nabla_{b}\psi=0.
\end{aligned}
\end{equation}
From the latter, projecting with respect to the timelike frame, one obtains
\begin{eqnarray}
&&N^{b}\left(\nabla^{a}T_{ab}\right)=0, \quad \pounds_{\text{\tiny{\bmath{N}}}}\psi+\chi\psi+\frac{d\mathcal{V}}{d\phi}=0,
\label{KGequation} \\
&& h_{c}^{\,\,\,b}\left(\nabla^{a}T_{ab}\right)=0, \,\,\,\quad\quad\quad D_{c}\psi+\psi a_{c}=0.
\label{momentumcons}
\end{eqnarray}
Moreover, using the fact that $D_{a}\phi=0$ in the orthogonal
subspaces to $\bmath{e}_{0}$, one obtains from equation \eqref{commutatorframes}
\begin{equation*}
\left[\bar{\bmath{e}}_{a},\bar{\bmath{e}}_{b}\right]\phi=2\,\left(\chi^{A}\right)_{ab}\psi=0,
\end{equation*}
which implies
\begin{equation}
\left(\chi^{A}\right)_{ab}=0.
\label{commutatorframes_scalarfield}
\end{equation}
\begin{Remark}
 Following Friedrich in \cite{FR00}, one
could as well have defined
\begin{equation}
\nabla^{a}T_{ab}=q_{b}+qN_{b}, \quad J_{ab}=\nabla_{[a}q_{b]}.
\label{Alternative}
\end{equation}
Then, instead of using the  condition on the vanishing of the divergence of the energy-momentum tensor, 
one could include the equations $q=0$ and $q_{b}=0$ as a part of the equations determining the Einstein-nonlinear 
scalar field system in the frame representation. Once the gauge is fixed, the first equation in \eqref{Alternative} 
appears in the reduced system of evolution equations while the second part is regarded as a {\it zero quantity}, see equation (4.44) in \cite{FR00}. It can be shown that the zero quantities satisfy a system of 
\emph{subsidiary evolution equations}. For this, it can be shown that the zero quantities vanish if they are 
zero on the initial hypersurface. For the quantity $q_{b}$, the relevant subsidiary equation is given in equation 
(4.70) of \cite{FR00}. We also notice that the evolution for the acceleration can be computed from the tensor 
$J_{ab}$.
\end{Remark}
%
%
\section{The Einstein-Friedrich-nonlinear scalar field system}
In this section, we derive a first order symmetric hyperbolic system
for the EFE coupled to a nonlinear scalar field. Making use of the
Bianchi identity and the energy-momentum tensor given by equation
\eqref{stress}, we derive the propagation equations for the
\emph{electric} and \emph{magnetic} parts of the conformal Weyl
tensor. After fixing the gauge, we complete the reduced system of
evolution equations by deriving equations for the frame and the
connection coefficients. In the last part of this section, we make some
remarks concerning the hyperbolicity of the system.
\subsection{Basic expressions}
We start by computing the various components for the Friedrich tensor
$\bmath{F}$. Using equations \eqref{DecompositionWeyl} and
(\ref{Shoutenscalarfield}), one finds
\begin{eqnarray}
\label{Fprimes}
  \bar{F}_{00c0}&=&0=-\bar{F}_{000c},\quad \bar{F}_{00cd}=0=-\bar{F}_{00dc}, \nonumber\\
  \bar{F}_{a00d}&=&-E_{ad}+\frac{1}{6}\left(\frac{5}{2}\psi^{2}-\mathcal{V}(\phi)\right)h_{ad}=-\bar{F}_{a0d0}, \nonumber\\
  \bar{F}_{ab0d}&=&B_{dp}\epsilon^{p}{}_{ab}=-\bar{F}_{abd0}=-\bar{F}_{ba0d}, \\
  \bar{F}_{0bcd}&=&B_{bp}\epsilon^{p}{}_{cd}=-\bar{F}_{0bdc}=-\bar{F}_{b0cd}, \nonumber\\
  \bar{F}_{0b0d}&=&E_{bd}+\frac{1}{6}\left(\frac{1}{2}\psi^{2}+\mathcal{V}(\phi)\right)h_{bd}=-\bar{F}_{0bd0,} \nonumber\\
  \bar{F}_{abcd}&=&-2\left(h_{b[c}E_{d]a}-h_{a[c}E_{d]b}\right)-\frac{1}{6}\left(\frac{1}{2}\psi^{2}+\mathcal{V}(\phi)\right)\left(h_{ac}h_{db}-h_{ad}h_{cb}\right),\nonumber
\end{eqnarray}
with the non-vanishing traces
\begin{eqnarray}
\label{Fcomponentstrace}
h^{ac}\bar{F}_{a0c0}&=&\frac{1}{2}\left(\frac{5}{2}\psi^{2}-\mathcal{V}(\phi)\right), \nonumber\\
h^{bd}\bar{F}_{0b0d}&=&\frac{1}{2}\left(\frac{1}{2}\psi^{2}+\mathcal{V}(\phi)\right)=-h^{bd}\bar{F}_{0bd0}, \\
h^{bd}\bar{F}_{abcd}&=&E_{ac}-\frac{1}{3}\left(\frac{1}{2}\psi^{2}+\mathcal{V}(\phi)\right)h_{ac}=\bar F^{b}{}_{abc},\nonumber\\
h^{ac}h^{bd}\bar{F}_{abcd}&=&-\frac{1}{2}\psi^{2}-\mathcal{V}(\phi).\nonumber
\end{eqnarray}
Using expression \eqref{Fdual}, with equations
\eqref{DecompositionWeylDual} and \eqref{Shoutenscalarfield}, we get
the following components of the dual ${^{\star}}\bmath{F}$:
\begin{eqnarray}
\label{Fdualcomponents}
{^{\star}}\bar{F}_{00c0}&=&-{^{\star}}\bar{F}_{000c}=0,\quad {^{\star}}\bar{F}_{00cd}=-{^{\star}}\bar{F}_{00dc}=0,\nonumber\\
{^{\star}}\bar{F}_{a0c0}&=&-{^{\star}}\bar{F}_{a00c}=B_{ac}, \nonumber\\
{^{\star}}\bar{F}_{abc0}&=&-2E_{p[b}\epsilon_{a]}{}^p{}_c-\frac{1}{6}\left(\frac{1}{2}\psi^{2}+\mathcal{V}(\phi)\right)\epsilon_{bac}=-{^{\star}}\bar{F}_{ab0c}, \\
{^{\star}}\bar{F}_{a0cd}&=&E_{ap}\epsilon^{p}{}_{cd}-\frac{1}{6}\left(\frac{5}{2}\psi^{2}-\mathcal{V}(\phi)\right)\epsilon_{acd}, \nonumber\\
{^{\star}}\bar{F}_{0bcd}&=&-E_{bp}\epsilon^{p}{}_{cd}-\frac{1}{6}\left(\frac{1}{2}\psi^{2}+\mathcal{V}(\phi)\right)\epsilon_{bcd}, \nonumber\\
{^{\star}}\bar{F}_{0b0d}&=&B_{bd}, \quad {^{\star}}\bar{F}_{abcd}=-B_{pq}\epsilon^{p}{}_{ab}\epsilon^{q}{}_{cd}.\nonumber
\end{eqnarray}

\subsection{The Bianchi equations}
If one substitutes the expressions for the Friedrich tensor derived in
the previous section into the first Bianchi identities
$(\ref{GeomDecBI})$, one obtains the following relations for the
components of the zero quantity $\tilde{F}_{abc}$:
\begin{small}
\begin{eqnarray}
\label{Fstuff}
 \bar{\tilde{F}}_{0c0}&=&-\pounds_{\text{\tiny{$\bmath{N}$}}}\bar{F}_{00c0}+D^{q}\bar{F}_{q0c0}+\chi_{c}^{\,\,\,s}\bar{F}_{00s0}-\chi \bar{F}_{00c0}-\chi^{qb}\left(\bar{F}_{qbc0}+\bar{F}_{q0cb}\right)+a^{b}\left(\bar{F}_{0bc0}+\bar{F}_{00cb}+\bar{F}_{b0c0}\right),\nonumber\\
 \bar{\tilde{F}}_{0cd}&=&-\pounds_{\text{\tiny{$\bmath{N}$}}}\bar{F}_{00cd}+D^{q}\bar{F}_{q0cd}+a^{b}\bar{F}_{0bcd}+a^{q}\bar{F}_{q0cd}-\chi^{qb}\bar{F}_{qbcd}-\chi \bar{F}_{00cd}-\chi^{q}_{\,\,\,c}\bar{F}_{q00d}-\chi^{q}_{\,\,\,d}\bar{F}_{q0c0} \nonumber\\
		 &&+\chi_{c}^{\,\,\,s}\bar{F}_{00sd}+\chi_{d}^{\,\,\,s}\bar{F}_{00cs}+a_{c}\bar{F}_{000d}+a_{d}\bar{F}_{00c0},\nonumber\\
  \bar{\tilde{F}}_{b0d}&=&-\pounds_{\text{\tiny{$\bmath{N}$}}}\bar{F}_{0b0d}+D^{a}\bar{F}_{ab0d}-\chi \bar{F}_{0b0d}-\chi^{a}_{\,\,\,b}\bar{F}_{a00d}-\chi^{ac}\bar{F}_{abcd}+\chi_{b}^{\,\,\,s}\bar{F}_{0s0d}+a_{b}\bar{F}_{000d}+\chi_{d}^{\,\,\,s}\bar{F}_{0b0s} \\
		   &&+a^{q}\bar{F}_{qb0d}+a^{c}\bar{F}_{0bcd},\nonumber\\
 \bar{\tilde F}_{bcd}&=&-\pounds_{\text{\tiny{$\bmath{N}$}}}\bar{F}_{0bcd}+D^{a}\bar{F}_{abcd}+a^{q}\bar{F}_{qbcd}+\left(a_{b}\bar{F}_{00cd}+a_{c}\bar{F}_{0b0d}+a_{d}\bar{F}_{0bc0}\right)-\chi \bar{F}_{0bcd} \nonumber\\
		  &&+\chi_{b}^{\,\,\,q}\bar{F}_{0qcd}+\chi_{c}^{\,\,\,q}\bar{F}_{0bqd}+\chi_{d}^{\,\,\,q}\bar{F}_{0bcq}-\chi^{q}_{\,\,\,b}\bar{F}_{q0cd}-\chi^{q}_{\,\,\,c}\bar{F}_{qb0d}-\chi^{q}_{\,\,\,d}\bar{F}_{qbc0},\nonumber
\end{eqnarray}
\end{small}
where we have used the fact that $\bmath{F}$ is anti-symmetric in the
last two indices, see e.g. \cite{Fri98}. Similar relations
hold for the dual ${^{\star}}\tilde{\bmath{F}}$.
\begin{Remark}
 In \cite{FR00} ---cfr. equation
(4.47)--- suitable zero quantities are defined by using the
decomposition in terms of irreducible components of
$\tilde{\bmath{F}}$.
\end{Remark}
\subsubsection{The evolution equation for the electric part of the Weyl tensor}
An evolution equation for the electric part of the Weyl tensor can be
obtained using the third equation of \eqref{Fstuff} together with
the expressions \eqref{Fprimes}-\eqref{Fdualcomponents}, and
then symmetrising with respect to the indices $(bd)$, giving
\begin{equation*}
\begin{aligned}
 \bar{\tilde{F}}_{(b|0|d)}=&-\pounds_{\text{\tiny{\bmath{N}}}}E_{bd}-\frac{1}{6}\left(\frac{1}{2}\psi^{2}+\mathcal{V}(\phi)\right)\pounds_{\text{\tiny{\bmath{N}}}}h_{bd}-\frac{1}{6}h_{bd}\pounds_{\text{\tiny{\bmath{N}}}}\left(\frac{1}{2}\psi^{2}+\mathcal{V}(\phi)\right)+D_{a}B_{p(d}\epsilon_{b)}{}^{pa} \\
		      &+2a_{a}B_{p(b}\epsilon_{d)}{}^{pa}-2\chi E_{bd}+2\chi^{a}{}_{(b}E_{d)q}+3\chi_{(b}{}^{q}E_{d)q}-h_{db}\chi^{ac}E_{ac}-\frac{1}{3}\left(\psi^{2}-\mathcal{V}(\phi)\right)\chi_{(bd)}.
\end{aligned}
\end{equation*}
Similarly, using equation \eqref{LieDer3metric}, we get
\begin{equation*}
 \begin{aligned}
  \bar{\tilde{F}}_{(b|0|d)}=&-\pounds_{\text{\tiny{\bmath{N}}}}E_{bd}+D_{a}B_{p(d}\epsilon_{b)}{}^{pa}+2a_{a}B_{p(b}\epsilon_{d)}{}^{pa}-2\chi E_{bd}+2\chi^{a}{}_{(b}E_{d)q}+3\chi_{(b}{}^{q}E_{d)q}-h_{db}\chi^{ac}E_{ac} \\
		       &-\frac{1}{2}\psi^{2}\chi_{(bd)}-\frac{1}{6}h_{bd}\pounds_{\text{\tiny{\bmath{N}}}}\left(\frac{1}{2}\psi^{2}+\mathcal{V}(\phi)\right).
\end{aligned}
\end{equation*}
The trace of the previous expression is given by
\begin{equation*}
 h^{rs}\bar{\tilde{F}}_{(r|0|s)}=-\frac{1}{2}\psi^{2}\chi-\frac{1}{2}\pounds_{\text{\tiny{\bmath{N}}}}\left(\frac{1}{2}\psi^{2}+\mathcal{V}(\phi)\right),
\end{equation*}
which is the evolution equation for the scalar field, i.e. the equation
expressing the conservation of energy. From this, it follows that $E_{ab}$ remains trace free during the evolution if
the data is given accordingly. 
Thus, taking the difference of the last two equations, and taking into account \eqref{commutatorframes_scalarfield}, 
the evolution equation for the components of the tensor $E_{ab}$ can be written as
\begin{eqnarray}
\label{evolutionE}
2 \pounds_{\text{\tiny{\bmath{N}}}}E_{bd}-2D_{a}B_{p(d}\epsilon_{b)}{}^{pa}&=&4a_{a}B_{p(b}\epsilon_{d)}{}^{pa}-4\chi E_{bd}+10\chi^{q}{}_{(b}E_{d)q}-2h_{db}\chi^{ac}E_{ac} \\
										&&-\psi^{2}\left(\chi_{(bd)}-\frac{1}{3}\chi h_{bd}\right).\nonumber
\end{eqnarray}

\subsubsection{The evolution equation for the magnetic part of the Weyl tensor}
An evolution equation for the magnetic part of the Weyl tensor can also be
derived from the analogue of the third equation of \eqref{Fstuff} for
the Hodge dual, using the expressions
\eqref{Fprimes}-\eqref{Fdualcomponents} to give
\begin{equation*}
 \begin{aligned}
  {^{\star}}\bar{\tilde{F}}_{b0d}=&-\pounds_{\text{\tiny{\bmath{N}}}}B_{bd}+D^{a}\left(2E_{p[b}\epsilon_{a]}{}^p{}_d +\frac{1}{6}\left(\frac{1}{2}\psi^{2}+\mathcal{V}(\phi)\right)\epsilon_{bad}\right)-\chi B_{bd}+\chi^{a}{}_{b}B_{ad}+2\chi_{(b}{}^{a}B_{d)a} \\
          &+2a_{q}B_{pb}\epsilon^{qp}{}_{d}+a^{q}E_{pq}\epsilon^{p}{}_{bd}+\chi^{qb}B_{pq}\epsilon^{p}{}_{ab}\epsilon^{q}{}_{cd}.
 \end{aligned}
\end{equation*}
Now, since $B_{bd}$ is a symmetric tensor, all the information about
its evolution is contained in the symmetrised expression of
${^{\star}}\tilde{F}_{(b|0|d)}$. Consequently, by symmetrising the previous
equation with respect to the spatial indices $(bd)$, and using \eqref{commutatorframes_scalarfield}, we get
\begin{equation}
\label{evolutionB}
2\pounds_{\text{\tiny{\bmath{N}}}}B_{bd}-2D_{a}E_{p(b}\epsilon_{d)}{}^{ap}=-4a_{a}E_{p(b}\epsilon_{d)}{}^{pa}+6\chi_{(b}{}^{a}B_{d)a}-2\chi B_{bd}+2\chi_{ac}B_{pq}\epsilon^{pa}{}_{(b}\epsilon_{d)}{}^{qc}.
\end{equation}
Ignoring the information about the trace, the principal part of the equations \eqref{evolutionE} and \eqref{evolutionB} 
is a symmetric matrix for the variables $E_{cd}$, $B_{cd}$, $c\leq d$, reading
\begin{equation*}
                    \begin{pmatrix}
			    2\bmath{e}_{0} & 0 & 0 & 0 & 0 & 0 & 0 & -D_{1} & D_{2} & D_{3} & -D_{3} & 0 \\
			    0 & 2\bmath{e}_{0} & 0 & 0 & 0 & 0 & D_{1} & 0 & -D_{3} & -D_{2} & 0 & D_{2} \\
		            0 & 0 & 2\bmath{e}_{0} & 0 & 0 & 0 & -D_{2} & D_{3} & 0 & 0 & D_{1} & -D_{1} \\
                            0 & 0 & 0 & \bmath{e}_{0} & 0 & 0 & -D_{3} & D_{2} & 0 & 0 & 0 & 0 \\
                            0 & 0 & 0 & 0 & \bmath{e}_{0} & 0 &D_{3} & 0 & -D_{1} & 0 & 0 & 0 \\
                            0 & 0 & 0 & 0 & 0 & \bmath{e}_{0} & 0 & -D_{2} & D_{1} & 0 & 0 & 0 \\
       		            0 & D_{1} & -D_{2} & -D_{3} & D_{3} & 0 & 2\bmath{e}_{0} & 0 & 0 & 0 & 0 & 0 \\
		            -D_{1} & 0 & D_{3} & D_{2} & 0 & -D_{2} & 0 & 2\bmath{e}_{0} & 0 & 0 & 0 & 0  \\
         		    D_{2} & -D_{3} & 0 & 0 & -D_{1} & D_{1} & 0 & 0 & 2\bmath{e}_{0} & 0 & 0 & 0 \\
                            D_{3} & -D_{2} & 0 & 0 & 0 & 0 & 0 & 0 & 0 & \bmath{e}_{0} & 0 & 0 \\
                            -D_{3} & 0 & D_{1} & 0 & 0 & 0 & 0 & 0 & 0 & 0 & \bmath{e}_{0} & 0 \\
                            0 & D_{2} & -D_{1} & 0 & 0 & 0 & 0 & 0 & 0 & 0 & 0 & \bmath{e}_{0}
		    \end{pmatrix}
                    \begin{pmatrix}
			    E_{12} \\
                            E_{13} \\
                            E_{23} \\
                            E_{11} \\
                            E_{22} \\
                            E_{33} \\
                            B_{12} \\
                            B_{13} \\
                            B_{23} \\
                            B_{11} \\
                            B_{22} \\
                            B_{33}
		    \end{pmatrix}.
\end{equation*}
\begin{Remark}
 The trace-freeness of the tensors $E_{ab}$ and
$B_{ab}$ can be recovered by assuming it initially. Then, using 
the evolution equations, it can be shown that $E_{ab}$ and
$B_{ab}$ are trace-free at later times (see e.g. the discussion in
\cite{Fri98} for the perfect fluid case).
\end{Remark}

\subsection{The Lagrangian description and  Fermi transport}
In order to deduce the remaining evolution equations, we will adopt a
Lagrangian description. This point of view amounts to requiring the
timelike vector of the orthonormal frame to follow the matter flow
lines. Accordingly, we introduce coordinates $(\bmath{x}, t)$ such that
\begin{equation}
 \bmath{e}_{0}=\partial_{t}, \quad e_{0}{}^{\mu}=\delta^{\mu}_{0}.
\label{lagrangiandescription}
\end{equation}
This particular choice is equivalent to setting
$\bmath{\theta}^{b}=\theta^{b}{}_{j}\mbox{d}x^{j}$ while, at the same
time, fixing the lapse function to one\footnote{See also
\cite{CBY01}, where a symmetric hyperbolic system was
obtained for the Einstein-Euler system. This construction holds for an
arbitrary Eulerian frame.}.  With this choice (since
$\pounds_{N}=\partial_{t}$), we have from equations 
(\ref{psi_{0}_frame}), (\ref{KGequation}), and
(\ref{constraint_Da_phi}) that
\begin{eqnarray}
&& \partial_{t}\phi=\psi \equiv N^{a}\psi_{a}=-\alpha<0,
\label{defipsi} \\
&& \partial_{t}\psi=-\psi\chi-\frac{d\mathcal{V}}{d\phi},
\label{evolutionpsi} \\
&&\bar{e}^{\,\,0}_{a}\psi=-\bar{e}^{\,\,j}_{a}\nabla_{j}\phi.
\label{D_phi} 
\end{eqnarray}
Now, the timelike coframe is given in terms of the natural cobasis
through the relation
\begin{equation*}
 \bmath{\theta}^{0}=\mbox{d}t+\beta_{j}\mbox{d}x^{j},
\end{equation*}
while the spatial frame vectors are found to be 
\begin{equation}\label{spatialframe_coeff}
 \bar{\bmath{e}}_{a}=(\theta_{a}{}^{j})^{-1}\left(\partial_{j}-\beta_{j}\partial_{t}\right), \quad \bar{e}_{a}{}^{0}=(\theta_{a}{}^{j})^{-1}\beta_{j}, \quad \bar{e}_{a}{}^{j}=(\theta_{a}{}^{j})^{-1}.
\end{equation}
It then follows, from equations \eqref{D_phi} and \eqref{spatialframe_coeff}, that 
\begin{equation}
 \beta_{j}=-\frac{1}{\psi}\partial_{j}\phi.
\label{Beta_j}
\end{equation}
Thus, since $\beta_{j}$ is nonzero, the surfaces of constant time are
not necessarily spacelike for the characteristic cone and this could
be a problem for the hyperbolicity of the system, see
\cite{CBY01,CBY02}. Finally, the
remaining frame components are chosen to be Fermi propagated along
$\bmath{e}_{0}$. That is, we require
\begin{equation*}
 \nabla_{0}\bmath{e}_{a}-\left(\mathbf{g}\left(\bmath{e}_{a},\nabla_{0}\bmath{e}_{0}\right)\bmath{e}_{0}-\mathbf{g}\left(\bmath{e}_{a},\bmath{e}_{0}\right)\nabla_{0}\bmath{e}_{0}\right)=0,
\end{equation*}
which implies
\begin{equation*}
 \bar{\gamma}^{a}{}_{b0}=0.
\end{equation*}

\subsection{Evolution equation for the frame coefficients}
As already mentioned, the evolution equations for the components of
the frame are obtained from the
relation \eqref{commutatorframes} which yields
\begin{equation*}
 \left[\bmath{e}_{0},\bar{\bmath{e}}_{b}\right]=a_{b}\bmath{e}_{0}-\bar{\gamma}^{c}{}_{0b}\bar{\bmath{e}}_{c},
\end{equation*}
where $\bar{\gamma}^{c}{}_{b0}=0$ (Fermi
gauge) has been used. Therefore, the evolution equations for the remaining frame
coefficients read
\begin{equation}\label{evolutionspatialframes}
\begin{aligned}
 \partial_{t}\bar{e}_{b}{}^{i}&=-\chi_{b}{}^{c}\bar{e}_{c}{}^{i}, \\
 \partial_{t}\bar{e}_{b}{}^{0}&=a_{b}-\chi_{b}{}^{c}\bar{e}_{c}{}^{0},
\end{aligned}
\end{equation}
which, together with equation \eqref{spatialframe_coeff}, imply propagation
equations for the components of the metric in the local coordinate
system.  In particular, one has
\begin{equation*}
 \partial_{t}\beta_{j}=\bar{\theta}^{b}{}_{j}a_{b},
\end{equation*}
with $\beta_{j}$ given by equation \eqref{Beta_j}, see also equation
(6.2) in \cite{CBY01} for an arbitrary lapse $U$. 

\subsection{Evolution equations for the connection coefficients}
The equations for the connection coefficients are obtained from the
splitting of the Riemann tensor with respect to the frame
$\{\bmath{e}_a\}$. In general, we have
\begin{equation}
\begin{aligned}
\bar{R}^{a}{}_{b0d}&=\bmath{e}_{0}\left(\gamma^{a}{}_{bd}\right)-D_{d}\gamma^{a}{}_{b0}-a_{d}\gamma^{a}{}_{b0}
                              -\left(\gamma^{p}{}_{d0}-\chi_{d}{}^{p}\right)\bar{\gamma}^{a}{}_{bp}-a_{b}\chi_{d}{}^{a}+a^{a}\chi_{db}, \\
\bar{R}^{0}{}_{b0d}&=\bmath{e}_{0}\left(\chi_{db}\right)-D_{d}a_{b}-a_{b}a_{d}+\chi_{pb}\chi_{d}{}^{p}-\chi_{dp}\gamma^{p}{}_{b0}-\chi_{pb}\gamma^{p}{}_{d0}, \\ 
\bar{R}^{a}{}_{0cd}&=D_{c}\chi_{d}{}^{a}-D_{d}\chi_{c}{}^{a}-a^{a}\left(\chi_{cd}-\chi_{dc}\right), \\
\bar{R}^{a}{}_{bcd}&=\,^{3}R^{a}{}_{bcd}+\chi_{c}{}^{a}\chi_{db}-\chi_{d}{}^{a}\chi_{cb}-\gamma^{a}{}_{b0}\left(\chi_{cd}-\chi_{dc}\right),
\end{aligned}
\label{Riemann_connection}
\end{equation}
where $^{3}R^{a}{}_{bcd}$ denotes the Riemann tensor constructed only
with the spatial connection coefficients $\bar{\gamma}^{c}{}_{ab}$.
The first two identities give evolution equations once the Lagrangian
gauge is introduced. The remaining two equations are the
\emph{quasi-constraints} for the connection coefficients (see
\cite{CBY01,CBY02}). No equations for the connection coefficient
associated to the acceleration can be deduced from these
identities. In the sequel, it will be shown how evolution equations
for the acceleration can be obtained for our particular problem.

From equations \eqref{Riemann_connection}, we can also deduce two
important equations relating the Ricci tensor to the connection:
\begin{equation}
 \begin{aligned}
  R_{00}&=-\bmath{e}_{0}\left(\chi\right)+D_{p}a^{p}-\chi_{p}{}^{b}\chi_{b}{}^{p}+a_{p}a^{p} \\
  \bar{R}_{0d}&=D_{c}\chi_{d}{}^{c}-D_{d}\chi-2\,a^{c}\left(\chi^{A}\right)_{cd}.
 \end{aligned}
\label{Ricci_connection}
\end{equation}
The first identity in \eqref{Riemann_connection}, together with the
conditions for the Lagrangian and Fermi gauge provide the equation
\begin{equation}
\partial_{t}\gamma^{a}{}_{bd}=-\gamma^{a}{}_{bp}\chi_{d}{}^{p}+2h^{ap}\chi_{d[p}a_{b]}+B_{dp}\epsilon^{pa}{}_{b},
\label{evolutionspatial_connectioncoefficients}
\end{equation}
describing the evolution of the spatial connection coefficients
$\bar{\gamma}^{c}{}_{ab}$. To obtain the last equation we have used $\bar{R}^{a}{}_{b0d}=\bar{C}^{a}{}_{b0d}=B_{dp}\epsilon^{pa}{}_{b}$. 

The evolution equation for the part of the connection described by
$\chi_{bd}$ is obtained from the second identity in
\eqref{Riemann_connection}. In order to do so, first, we will derive
the evolution and the quasi-constraint equations for the
acceleration. The evolution equation for the acceleration can be
obtained from
\begin{equation*}
\begin{aligned}
 \left[\bmath{e}_{0},\bar{\bmath{e}}_{c}\right]\psi&=c^{0}{}_{0c}\bmath{e}_{0}(\psi)+c^{p}{}_{0c}\bar{\bmath{e}}{}_{p}(\psi) \\
                                                      &=\gamma^{0}{}_{c0}(\partial_{t}\psi)+\left(\gamma^{p}{}_{c0}-\gamma^{p}{}_{0c}\right)(D_{p}\psi) \\
                                                      &=a_{c}\left(\partial_{t}\psi\right)-\chi_{c}{}^{p}\left(D_{p}\psi\right)  ,
\end{aligned}
\end{equation*}
where the properties of the Lagrangian and Fermi gauge have been
employed. Now, expanding the left hand side and making use of the
evolution and the quasi-constraint equation for the energy-momentum
tensor of the scalar field, one has
\begin{equation*}
 \partial_{t}a_{c}-D_{c}\chi=-\chi_{c}^{\,\,\,p}a_{p}+\left(\chi+\frac{2}{\psi}\frac{d\mathcal{V}}{d\phi}\right)a_{c}
\end{equation*}
so that, using the second equation in \eqref{Ricci_connection}, we arrive at
\begin{equation}
 \partial_{t}a_{c}-D_{p}\chi_{c}{}^{p}=-2\left(\chi^{A}\right)_{pc}a^{p}-\chi_{c}{}^{p}a_{p}+\left(\chi+\frac{2}{\psi}\frac{d\mathcal{V}}{d\phi}\right)a_{c}.
\label{evolution_a}
\end{equation}
In the case of the quasi-constraint, a computation yields
\begin{equation*}
D_{c}a_{b}-D_{b}a_{c}=2\left(\chi+\frac{1}{\psi}\frac{d\mathcal{V}}{d\phi}\right)\left(\chi^{A}\right)_{cb}.
\end{equation*}
Thus, making use of this equation in the second identity of \eqref{Riemann_connection} and recalling the properties of the Fermi
gauge, one finds 
\begin{equation}
\partial_{t}\chi_{db}-D_{b}a_{d}=-E_{db}+\frac{1}{3}\left(\mathcal{V}(\phi)-\psi^{2}\right)h_{db}-\chi_{d}{}^{p}\chi_{pb}+2\left(\chi+\frac{1}{\psi}\frac{d\mathcal{V}}{d\phi}\right)\left(\chi^{A}\right)_{db}+a_{d}a_{b}.
\label{evolution_chi}
\end{equation}
The principal part of the combined system of equations \eqref{evolution_a} and \eqref{evolution_chi} is given by 
\begin{equation*}
 \begin{pmatrix}
  \bmath{e}_{0} & -D_{1} & -D_{2} & -D_{3} \\
  -D_{1} & \bmath{e}_{0} & 0 & 0 \\
  -D_{2} & 0 & \bmath{e}_{0} & 0 \\
  -D_{3} & 0 & 0 & \bmath{e}_{0}
 \end{pmatrix}
 \begin{pmatrix}
  a_{d} \\
  \chi_{d}{}^{1} \\
  \chi_{d}{}^{2} \\
  \chi_{d}{}^{3} 
 \end{pmatrix}.
\end{equation*}
which is symmetric. Finally, since for our particular problem one has
$(\chi^A)_{ab}=0$,  equation \eqref{evolution_a} takes the form
\begin{equation}
\partial_{t}a_{c}-D_{p}\chi_{c}{}^{p}=\left(\frac{2}{\psi}\frac{d\mathcal{V}}{d\phi}+\chi\right)a_{c}-\chi_{c}{}^{p}a_{p},
\label{evolutionA}
\end{equation}
and, after
symmetrising \eqref{evolution_chi}, we obtain
\begin{equation}
2\partial_{t}\chi_{(bd)}-2D_{(b}a_{d)}
				    =\frac{2}{3}\left(\mathcal{V}(\phi)-\psi^{2}\right)h_{bd}-2\chi^{p}_{\,\,(d}\chi_{b)p}+2a_{b}a_{d}-2E_{bd}.	    
\label{evolutionsecondfundform}
\end{equation}
Also, from the first equation in \eqref{Ricci_connection}, it follows that
\begin{equation}\label{Evol_trace_2ndFF}
 \partial_{t}\chi-D_{p}a^{p}=\mathcal{V}(\phi)-\psi^{2}-\chi^{cd}\chi_{cd}+a^{2},
\end{equation}
where $a^{2}=a_{c}a^{c}$.
Then,  the principal part of the sytem reads

\begin{equation*}
 \begin{pmatrix}
  \bmath{e}_{0} & 0 & 0 & -D_{2} & -D_{3} & 0 & -D_{1} & 0 & 0 \\
  0 & \bmath{e}_{0} & 0 & -D_{1} & 0 & -D_{3} & 0 & -D_{2} & 0 \\
  0 & 0 & \bmath{e}_{0} & 0 & -D_{1} & -D_{2} & 0 & 0 & -D_{3} \\
  -D_{2} & -D_{1} & 0 & 2\bmath{e}_{0} & 0 & 0 & 0 & 0 & 0 \\
  -D_{3} & 0 & -D_{1} & 0 & 2\bmath{e}_{0} & 0 & 0 & 0 & 0 \\
  0 & -D_{3} & -D_{2} & 0 & 0 & 2\bmath{e}_{0} & 0 & 0 & 0 \\
  -D_{1} & 0 & 0 & 0 & 0 & 0 & \bmath{e}_{0} & 0 & 0 \\
   0 & -D_{2} & 0 & 0 & 0 & 0 & 0 & \bmath{e}_{0} & 0 \\
  0 & 0 & -D_{3} & 0 & 0 & 0 & 0 & 0 & \bmath{e}_{0}
 \end{pmatrix}
 \begin{pmatrix}
  a_{1} \\
  a_{2} \\
  a_{3} \\
  \chi_{12} \\
  \chi_{13} \\
  \chi_{23} \\
  \chi_{11} \\
  \chi_{22} \\
  \chi_{33} \\ 
 \end{pmatrix}.
\end{equation*}
which is clearly symmetric.

\subsection{Hyperbolicity considerations}
The system consisting of equations \eqref{evolutionE},
\eqref{evolutionB}, \eqref{defipsi}, \eqref{evolutionpsi},
\eqref{evolutionspatialframes},
\eqref{evolutionspatial_connectioncoefficients}, \eqref{evolutionA}
and \eqref{evolutionsecondfundform} can be written matricially as
\begin{equation}
 \mathbf{A}^{0}\partial_{t}\mathbf{u}-\mathbf{A}^{p}\bar{\bmath{e}}_{p}(\mathbf{u})=\mathbf{B}(\mathbf{u})\mathbf{u}.
\label{framederiv_syst}
\end{equation}
As discussed in \cite{CBY01}, these systems are not hyperbolic in the
usual sense as, in general, the time lines are not hypersurface
orthogonal and the ``spatial'' frame vectors $\bar{\bmath{e}}_{a}$
have components in the time direction - cfr. equation
\eqref{spatialframe_coeff}. Since the surfaces of constant time $t$
are not necessarily spacelike, this type of system is referred to as a
{\it quasi-FOSH} system \cite{CBY01}. In terms of the partial
derivatives, equation \eqref{framederiv_syst} reads
\begin{equation}
 \tilde{\mathbf{A}}^{0}(\mathbf{u})\partial_{t}\mathbf{u}-\mathbf{A}^{j}(\mathbf{u})\partial_{j}\mathbf{u}=\mathbf{B}(\mathbf{u})\mathbf{u}
\label{partialderiv_syst}
\end{equation}
with
\begin{equation}
\tilde{\mathbf{A}}^{0}(\mathbf{u})\equiv
\mathbf{A}^{0}-\mathbf{A}^{p}\bar{e}_{p}{}^{0}, \quad \mathbf{A}^{j}(\mathbf{u})\equiv\mathbf{A}^{p}\bar{e}_{p}{}^{j}. 
\label{A0_scalarfield}
\end{equation}
In order to have a well posed initial value problem, the matrix $\tilde{\mathbf{A}}^{0}(\mathbf{u})$ must be positive definite. 
This is the case, as long the quadratic form
\begin{equation}
\sum_{b=1,2,3}\theta^{b}{}_{i}\theta^{b}{}_{j} -\beta_{i}\beta_{j}
\label{positive_quadratic_form}
\end{equation}
is positive definite, see Proposition 9 in
\cite{CBY01}.  In the next section, we will consider a
reference solution admitting a foliation by homogeneous spacelike
hypersurfaces. As a consequence,  the linearisation of the system \eqref{nonlinearequation} is well posed 
without the need to control the smallness of $\beta_{i}$. The
smallness of these terms will be taken care by the perturbation fields, see also \cite{Reu99}.

Written in terms of partial derivatives, our system of evolution equations reads
\begin{small}\begin{equation}\label{EFsf}
 \begin{aligned}
 \partial_{t}\phi&=\psi, \\ 
 \partial_{t}\psi&=-\psi\chi-\frac{d\mathcal{V}}{d\phi}, \\
 2\partial_{t}\chi_{(bd)}-2\,\bar{e}_{(b}{}^{0}\partial_{t}a_{d)}-2\,\bar{e}_{(b}{}^{j}\partial_{j}a_{d)}&=\frac{2}{3}\left(\mathcal{V}(\phi)-\psi^{2}\right)h_{bd}-2\chi^{p}_{\,\,(d}\chi_{b)p}+2a_{b}a_{d}-2E_{bd} \\
                 &\quad-\left(\gamma^{p}{}_{bd}+\gamma^{p}{}_{db}\right)a_{p}, \\
 \partial_{t}a_{c}-\bar{e}_{p}{}^{0}\partial_{t}\chi_{c}{}^{p}-\bar{e}_{p}{}^{j}\partial_{j}\chi_{c}{}^{p}&=\left(\frac{2}{\psi}\frac{d\mathcal{V}}{d\phi}+\chi\right)a_{c}-\chi_{c}{}^{p}a_{p}-\gamma^{q}{}_{cp}\chi{}_{q}{}^{p}+\gamma^{p}{}_{qp}\chi_{c}{}^{q}, \\
 2\partial_{t}E_{bd}-2\epsilon^{pa}{}_{(b|}\bar{e}_{a}{}^0\partial_{t}B_{p|d)}-2\epsilon^{pa}{}_{(b|}\bar{e}_{a}{}^j\partial_{j}B_{p|d)}&=-\psi^{2}\left(\chi_{(bd)}-\frac{1}{3}\chi h_{bd}\right)-4\chi E_{bd}+10\chi^{q}_{\,\,(b}E_{d)q} \\
                                                                                                                                              &\quad-2h_{bd}\chi^{qp}E_{qp}+4a_{a}B_{p(b}\epsilon_{d)}{}^{pa}-2\gamma^{q}{}_{pa}B_{q(d}\epsilon_{b)}{}^{pa} \\
                                                                                                                                              &\quad-2\epsilon^{pa}{}_{(d}\gamma^{q}{}_{b)a}B_{pq}, \\  
 2\partial_{t}B_{bd}-2\epsilon^{ap}{}_{(d|}\bar{e}_{a}{}^{0}\partial_{t}E_{|b)p}-2\epsilon^{ap}{}_{(d}\bar{e}_{|a}{}^{j}\partial_{j|}E_{b)p}&=-2\chi B_{bd}+6\chi^{q}_{\,\,(b}B_{d)q}+2\chi_{ac}B_{pq}\epsilon^{pa}{}_{(b}\epsilon_{d)}{}^{qc}\\
									   &\quad-4a_{a}E_{p(b}\epsilon_{d)}{}^{pa}-2\gamma^{q}{}_{pa}E_{q(b}\epsilon_{d)}{}^{ap} \\
                                                                           &\quad-2\epsilon^{ap}{}_{(b}\gamma^{q}{}_{d)a}E_{pq}, \\
 \partial_{t}\bar{\gamma}^{a}{}_{bd}&=B_{dp}\epsilon^{pa}{}_{b}-\chi_{d}{}^{p}\bar{\gamma}^{a}{}_{bp}+2h^{ap}\chi{}_{d[p}a_{b]}, \\ 
 \partial_{t}\bar{e}_{b}{}^{i}&=-\chi_{b}{}^{c}\bar{e}_{c}{}^{i},\\
 \partial_{t}\bar{e}_{b}{}^{0}&=-\chi_{b}{}^{c}\bar{e}_{c}{}^{0}+a_{b}.
 \end{aligned}
\end{equation}              \end{small}
This system has the form given by equation \eqref{partialderiv_syst}. If one writes 
\begin{equation}
 \mathbf{u}^{T}=\left(\phi,\,\,\psi,\,\,\mathbf{z}^{T},\,\,\mathbf{w}^{T},\,\,\mathbf{x}^{T},\,\,\mathbf{y}^{T}\right),
\label{u}
\end{equation}
where
\begin{equation*}
\begin{aligned}
 \mathbf{z}^{T}&=\left(\chi_{11},\,\,\chi_{22},\,\,\chi_{33},\,\,\chi_{12},\,\,\chi_{13},\,\,\chi_{23},\,\,a_{1},\,\,a_{2},\,\,a_{3}\right), \\
 \mathbf{w}^{T}&=\left(E_{12},\,\,E_{13},\,\,E_{23},\,\,E_{11},\,\,E_{22},\,\,E_{33},\,\,B_{12},\,\,B_{13},\,\,B_{23},\,\,B_{11},\,\,B_{22},\,\,B_{33}\right), \\
 \mathbf{x}^{T}&=\left(e_{1}{}^{0},\,\,e_{2}{}^{0},\,\,e_{3}{}^{0},\,\,e_{1}{}^{1},\,\,e_{1}{}^{2},\,\,e_{1}{}^{3},\,\,e_{2}{}^{1},\,\,e_{2}{}^{2},\,\,e_{2}{}^{3},\,\,e_{3}{}^{1},\,\,e_{3}{}^{2},\,\,e_{3}{}^{3}\right),  \\
 \mathbf{y}^{T}&=\left(\gamma^{1}{}_{22},\,\,\gamma^{1}{}_{33},\,\,\gamma^{1}{}_{23},\,\,\gamma^{2}{}_{11},\,\,\gamma^{2}{}_{33},\,\,\gamma^{2}{}_{31},\,\,\gamma^{3}{}_{11},\,\,\gamma^{3}{}_{22},\,\,\gamma^{3}{}_{12}\right), 
\end{aligned}
\end{equation*}
then the matrices given in \eqref{partialderiv_syst} and \eqref{A0_scalarfield} have the explicit form
\begin{equation}
 \tilde{\mathbf{A}}^{0}(\mathbf{u})=\begin{pmatrix}
			      \mathbf{I}_{2\times2} & 0 & 0 & 0\\
			      0 & \tilde{\mathbf{A}}^{0}_{9\times9} & 0 & 0 \\
                              0 & 0 &  \tilde{\mathbf{A}}^{0}_{12\times12} & 0 \\
                              0 & 0 & 0 & \mathbf{I}_{21\times21} 
			    \end{pmatrix}, \quad \quad
\mathbf{A}^{j}(\mathbf{u})=\begin{pmatrix}
			      \mathbf{0}_{2\times2} & 0 & 0 & 0\\
			      0 & \mathbf{A}^{j}_{9\times9} & 0 & 0 \\
                              0 & 0 &  \mathbf{A}^{j}_{12\times12} & 0 \\
                              0 & 0 & 0 & \mathbf{0}_{21\times21}
			   \end{pmatrix},
\label{A_matrices}
\end{equation}
with
\begin{equation*}
\tilde{\mathbf{A}}^{0}_{9\times9}=\begin{pmatrix}
	             1 & 0 & 0 & 0 & 0 & 0 & -e_{1}^{\,\,\,0} & 0 & 0 \\
		     0 & 1 & 0 & 0 & 0 & 0 & 0 & -e_{2}^{\,\,\,0} & 0 \\
		     0 & 0 & 1 & 0 & 0 & 0 & 0 & 0 & -e_{3}^{\,\,\,0} \\
		     0 & 0 & 0 & 2 & 0 & 0 & -e_{2}^{\,\,\,0} & -e_{1}^{\,\,\,0} & 0 \\
		     0 & 0 & 0 & 0 & 2 & 0 & -e_{3}^{\,\,\,0} & 0 & -e_{1}^{\,\,\,0} \\
		     0 & 0 & 0 & 0 & 0 & 2 & 0 & -e_{3}^{\,\,\,0} & -e_{2}^{\,\,\,0} \\
		     -e_{1}^{\,\,\,0} & 0 & 0 &  -e_{2}^{\,\,\,0} & -e_{3}^{\,\,\,0} & 0 &  1 & 0 & 0 \\
		     0 & -e_{2}^{\,\,\,0} & 0 &  -e_{1}^{\,\,\,0} & 0 & -e_{3}^{\,\,\,0} & 0 & 1 & 0 \\
		     0 & 0 & -e_{3}^{\,\,\,0} & 0 & -e_{1}^{\,\,\,0} & -e_{2}^{\,\,\,0} &  0 & 0 & 1 \\
		    \end{pmatrix},
\end{equation*}

\begin{equation*}
\tilde{\mathbf{A}}^{0}_{12\times12}=\begin{pmatrix}
			    2 & 0 & 0 & 0 & 0 & 0 & 0 & -e_{1}^{\,\,\,0} & e_{2}^{\,\,\,0} & e_{3}^{\,\,\,0} & -e_{3}^{\,\,\,0} & 0 \\
			    0 & 2 & 0 & 0 & 0 & 0 & e_{1}^{\,\,\,0} & 0 & -e_{3}^{\,\,\,0} & -e_{2}^{\,\,\,0} & 0 & e_{2}^{\,\,\,0} \\
		            0 & 0 & 2 & 0 & 0 & 0 & -e_{2}^{\,\,\,0} & e_{3}^{\,\,\,0} & 0 & 0 & e_{1}^{\,\,\,0} & -e_{1}^{\,\,\,0} \\
                            0 & 0 & 0 & 1 & 0 & 0 & -e_{3}^{\,\,\,0} & e_{2}^{\,\,\,0} & 0 & 0 & 0 & 0 \\
                            0 & 0 & 0 & 0 & 1 & 0 & e_{3}^{\,\,\,0} & 0 & -e_{1}^{\,\,\,0} & 0 & 0 & 0 \\
                            0 & 0 & 0 & 0 & 0 & 1 & 0 & -e_{2}^{\,\,\,0} & e_{1}^{\,\,\,0} & 0 & 0 & 0 \\
       		            0 & e_{1}^{\,\,\,0} & -e_{2}^{\,\,\,0} & -e_{3}^{\,\,\,0} & e_{3}^{\,\,\,0} & 0 & 2 & 0 & 0 & 0 & 0 & 0 \\
		            -e_{1}^{\,\,\,0} & 0 & e_{3}^{\,\,\,0} & e_{2}^{\,\,\,0} & 0 & -e_{2}^{\,\,\,0} & 0 & 2 & 0 & 0 & 0 & 0  \\
         		    e_{2}^{\,\,\,0} & -e_{3}^{\,\,\,0} & 0 & 0 & -e_{1}^{\,\,\,0} & e_{1}^{\,\,\,0} & 0 & 0 & 2 & 0 & 0 & 0 \\
                            e_{3}^{\,\,\,0} & -e_{2}^{\,\,\,0} & 0 & 0 & 0 & 0 & 0 & 0 & 0 & 1 & 0 & 0 \\
                            -e_{3}^{\,\,\,0} & 0 & e_{1}^{\,\,\,0} & 0 & 0 & 0 & 0 & 0 & 0 & 0 & 1 & 0 \\
                            0 & e_{2}^{\,\,\,0} & -e_{1}^{\,\,\,0} & 0 & 0 & 0 & 0 & 0 & 0 & 0 & 0 & 1
		    \end{pmatrix},
\end{equation*}

\begin{equation*}
\mathbf{A}^{j}_{9\times9}=\begin{pmatrix}
	             0 & 0 & 0 & 0 & 0 & 0 & e_{1}^{\,\,\,j} & 0 & 0 \\
		     0 & 0 & 0 & 0 & 0 & 0 & 0 & e_{2}^{\,\,\,j} & 0 \\
		     0 & 0 & 0 & 0 & 0 & 0 & 0 & 0 & e_{3}^{\,\,\,j} \\
		     0 & 0 & 0 & 0 & 0 & 0 & e_{2}^{\,\,\,j} & e_{1}^{\,\,\,j} & 0 \\
		     0 & 0 & 0 & 0 & 0 & 0 & e_{3}^{\,\,\,j} & 0 & e_{1}^{\,\,\,j} \\
		     0 & 0 & 0 & 0 & 0 & 0 & 0 & e_{3}^{\,\,\,j} & e_{2}^{\,\,\,j} \\
		     e_{1}^{\,\,\,j} & 0 & 0 &  e_{2}^{\,\,\,j} & e_{3}^{\,\,\,j} & 0 &  0 & 0 & 0 \\
		     0 & e_{2}^{\,\,\,j} & 0 &  e_{1}^{\,\,\,j} & 0 & e_{3}^{\,\,\,j} & 0 & 0 & 0 \\
		     0 & 0 & e_{3}^{\,\,\,j} & 0 & e_{1}^{\,\,\,j} & e_{2}^{\,\,\,j} &  0 & 0 & 0 \\
		    \end{pmatrix},
\end{equation*}
and
\begin{equation*}
\mathbf{A}^{j}_{12\times12}=\begin{pmatrix}
			    0 & 0 & 0 & 0 & 0 & 0 & 0 & e_{1}^{\,\,\,j} & -e_{2}^{\,\,\,j} & -e_{3}^{\,\,\,j} & e_{3}^{\,\,\,j} & 0 \\
			    0 & 0 & 0 & 0 & 0 & 0 & -e_{1}^{\,\,\,j} & 0 & e_{3}^{\,\,\,j} & e_{2}^{\,\,\,j} & 0 & -e_{2}^{\,\,\,j} \\
		            0 & 0 & 0 & 0 & 0 & 0 & e_{2}^{\,\,\,j} & -e_{3}^{\,\,\,j} & 0 & 0 & -e_{1}^{\,\,\,j} & e_{1}^{\,\,\,j} \\
                            0 & 0 & 0 & 0 & 0 & 0 & e_{3}^{\,\,\,j} & -e_{2}^{\,\,\,j} & 0 & 0 & 0 & 0 \\
                            0 & 0 & 0 & 0 & 0 & 0 & -e_{3}^{\,\,\,j} & 0 & e_{1}^{\,\,\,j} & 0 & 0 & 0 \\
                            0 & 0 & 0 & 0 & 0 & 0 & 0 & e_{2}^{\,\,\,j} & -e_{1}^{\,\,\,j} & 0 & 0 & 0 \\
       		            0 & -e_{1}^{\,\,\,j} & e_{2}^{\,\,\,j} & e_{3}^{\,\,\,j} & -e_{3}^{\,\,\,j} & 0 & 0 & 0 & 0 & 0 & 0 & 0 \\
		            e_{1}^{\,\,\,j} & 0 & -e_{3}^{\,\,\,j} & -e_{2}^{\,\,\,j} & 0 & e_{2}^{\,\,\,j} & 0 & 0 & 0 & 0 & 0 & 0  \\
         		    -e_{2}^{\,\,\,j} & e_{3}^{\,\,\,j} & 0 & 0 & e_{1}^{\,\,\,j} & -e_{1}^{\,\,\,j} & 0 & 0 & 0 & 0 & 0 & 0 \\
                            -e_{3}^{\,\,\,j} & e_{2}^{\,\,\,j} & 0 & 0 & 0 & 0 & 0 & 0 & 0 & 0 & 0 & 0 \\
                            e_{3}^{\,\,\,j} & 0 & -e_{1}^{\,\,\,j} & 0 & 0 & 0 & 0 & 0 & 0 & 0 & 0 & 0 \\
                            0 & -e_{2}^{\,\,\,j} & e_{1}^{\,\,\,j} & 0 & 0 & 0 & 0 & 0 & 0 & 0 & 0 & 0
		    \end{pmatrix}.
\end{equation*}
It can, therefore, be verified that $\tilde{\mathbf{A}}^{0}(\mathbf{u})$, $\mathbf{A}^{j}(\mathbf{u})$
are symmetric and, furthermore, that $\tilde{\mathbf{A}}^{0}(\mathbf{u})$ is positive definite as long 
as \eqref{positive_quadratic_form} is positive definite.

Using the standard theory of symmetric hyperbolic systems, one can then
conclude the local existence in time and uniqueness of smooth solutions
for the {\em evolution equations} implied by the Einstein-scalar field system. 
In order to conclude the existence of solutions to the {\em full}
Einstein-scalar field system, we note that, it follows from general arguments, that the constraint
equations are satisfied during the evolution, if they hold initially, see e.g.  \cite{FR00,Reu98}.
\begin{Remark}
\label{remark5}
 As part of the procedure to express the system
\eqref{EFsf} in an explicit symmetric hyperbolic form, one has to divide by $\psi^{2}$ the evolution equation \eqref{evolutionA}
for the acceleration. This could
imply that the system is not well behaved when $\psi\rightarrow0$. An inspection
shows that the potentially troublesome term is the one containing the first derivative of the 
potential, which,  by virtue of equations \eqref{defipsi}-\eqref{evolutionpsi} must be zero in this limit 
(possibly at $t\rightarrow+\infty$). Thus, we must require the coefficient $\mathcal{V}^{\prime}/\psi$ to be finite.
\end{Remark}
We summarise the results of this section as follows:
\begin{thm}
\label{Theorem:wellposedness}
The Einstein-Friedrich-nonlinear scalar field (EFsf) system
consisting of  the equations in \eqref{EFsf} 
forms a quasi-linear first-order symmetric
hyperbolic (FOSH) system for the scalar field, its momentum-density,
the frame coefficients, the connection coefficients and the electric
and magnetic parts of the Weyl tensor, relative to the slices of constant time $t$, 
as long as the quadratic form
\begin{equation*}
\sum_{a=1,2,3}\theta^{a}{}_{i}\theta^{a}{}_{j}-\frac{\partial_{i}\phi}{\psi}\frac{\partial_{j}\phi}{\psi},
\end{equation*}
is positive definite. Then, the local existence in time and uniqueness of smooth solutions is guaranteed.
\end{thm}
%
\section{Stability Analysis}
In this section, we use the symmetric hyperbolic system derived in
last section to show that, for some classes of potentials, the
evolution of sufficiently small nonlinear perturbations of a FLRW-nonlinear scalar field background,
prescribed on a Cauchy hypersurface with the topology of a 3-torus $\mathbb{T}^{3}$, 
have an asymptotic exponential decay. 

\subsection{New variables}
In order to simplify the analysis, we shall introduce new variables which will allow us to decouple the 
tracefree part of the second fundamental form as an independent variable. 
First, we introduce
\begin{equation}
\label{hubble}
 H\equiv \frac{\chi}{3},
\end{equation}
where $H$ is usually called the {\it Hubble function}. Then, since for our particular problem $(\chi^{A})_{bd}=0$, 
we write
\begin{equation}\label{Decomposition_2nFF}
 \chi_{(bd)}=\left(\chi^{ST}\right)_{bd}+Hh_{bd},
\end{equation}
where the evolution equation for $H$ is given by \eqref{Evol_trace_2ndFF} and reads now
\begin{equation*}
 \begin{aligned}
  3\partial_{t}H-D_{p}a^{p}=&-3H^{2}+\mathcal{V}(\phi)-\psi^{2}-\left(\chi^{ST}\right)_{bd}\left(\chi^{ST}\right)^{bd}+a_{p}a^{p}, 
\end{aligned}
\end{equation*}
while the evolution equation for $(\chi^{ST})_{bd}$ is obtained using \eqref{Decomposition_2nFF} in \eqref{evolutionsecondfundform} and then 
subtracting its trace \eqref{Evol_trace_2ndFF}, giving
\begin{equation*}
\begin{aligned}
2\partial_{t}\left(\chi^{ST}\right)_{db}-2\left(D_{(b}a_{d)}-\frac{h_{bd}}{3}D_{p}a^{p}\right)=&-4H\left(\chi^{ST}\right)_{db}-2\left(\chi^{ST}\right)_{pb}\left(\chi^{ST}\right)_{d}{}^{p}
+2a_{(b}a_{d)} \\
                     &+\frac{2}{3}\left[2\left(\chi^{ST}\right)^{2}-a^{2}\right]h_{bd} -2E_{bd}.
\end{aligned}
\end{equation*}
In turn, equation \eqref{evolution_a} reads
\begin{equation*}
 \partial_{t}a_{c}-D_{p}\left(\chi^{ST}\right)_{c}{}^{p}-D_{c}H=-Ha_{c}-\left(\chi^{ST}\right)_{c}^{\,\,\,p}a_{p}+\left(3H+\frac{2}{\psi}\frac{d\mathcal{V}}{d\phi}\right)a_{c}.
\end{equation*}
The block of the principal part of the system, where the above
decomposition is applied reads now
\begin{equation*}
 \begin{pmatrix}
  \bmath{e}_{0} & 0 & 0 & -D_{2} & -D_{3} & 0 & 0 & D_1 & D_1 & -D_{1}  \\
  0 & \bmath{e}_{0} & 0 & -D_{1} & 0 & -D_{3} & D_2 & 0 & D_2 & -D_{2} \\
  0 & 0 & \bmath{e}_{0} & 0 & -D_{1} & -D_{2} & D_3 & D_3 & 0 & -D_{3}\\
  -D_{2} & -D_{1} & 0 & 2\bmath{e}_{0} & 0 & 0 & 0 & 0 & 0 & 0 \\
  -D_{3} & 0 & -D_{1} & 0 & 2\bmath{e}_{0} & 0 & 0 & 0 & 0 & 0  \\
  0 & -D_{3} & -D_{2} & 0 & 0 & 2\bmath{e}_{0} & 0 & 0 & 0 & 0 \\
  -2D_{1} & D_{2} & D_{3} & 0 & 0 & 0 & 3\bmath{e}_{0} & 0 & 0 & 0\\
  D_{1} & -2D_{2} & D_{3} & 0 & 0 & 0 & 0 & 3\bmath{e}_{0} & 0 & 0 \\
  D_{1} & D_{2} & -2D_{3} & 0 & 0 & 0 & 0 & 0 & 3\bmath{e}_{0} & 0 \\
 -D_{1} & -D_{2} & -D_{3} & 0 & 0 & 0 & 0 & 0 & 0 & 3\bmath{e}_{0}
\end{pmatrix}
 \begin{pmatrix}
  a_{1} \\
  a_{2} \\
  a_{3} \\
  \left(\chi^{ST}\right)_{12} \\
  \left(\chi^{ST}\right)_{13} \\
  \left(\chi^{ST}\right)_{23} \\
  \left(\chi^{ST}\right)_{11} \\
  \left(\chi^{ST}\right)_{22} \\
  \left(\chi^{ST}\right)_{33} \\ 
  H
 \end{pmatrix},
\end{equation*}
which is non-symmetric. In order to recover the symmetry of the block,
we define six new variables
$\chi_{\pm}$, $E_{\pm}$ and $B_{\pm}$ via  (see e.g. \cite{FR00,WL05} for a similar
context)
\begin{subequations}
\begin{align}
 \chi_{+}\equiv\frac{1}{2}\left(\chi^{ST}_{22}+\chi^{ST}_{33}\right),\quad&\quad \chi_{-}\equiv\frac{1}{6}\left(\chi^{ST}_{22}-\chi^{ST}_{33}\right), \\
  E_{+}\equiv\frac{3}{2}\left(E_{22}+E_{33}\right),\quad&\quad  E_{-}\equiv\frac{1}{2}\left(E_{22}-E_{33}\right), \\
  B_{+}\equiv\frac{3}{2}\left(B_{22}+B_{33}\right),\quad&\quad  B_{-}\equiv\frac{1}{2}\left(B_{22}-B_{33}\right), 
\end{align}\label{NEWVAR}
\end{subequations}
and use the tracefree condition. It follows that 
\begin{equation}
\chi^{ST}_{11}=-2\chi_{+},\quad\quad\chi^{ST}_{22}=\chi_{+}+3\chi_{-},\quad\quad\chi^{ST}_{33}=\chi_{+}-3\chi_{-},
\end{equation}
\begin{equation}
E_{11}=-\frac{2}{3}E_{+},\quad\quad E_{22}=\frac{1}{3}E_{+}+E_{-},\quad\quad E_{33}=\frac{1}{3}E_{+}-E_{-},
\end{equation}
\begin{equation}
B_{11}=-\frac{2}{3}B_{+},\quad\quad B_{22}=\frac{1}{3}B_{+}+B_{-},\quad\quad B_{33}=\frac{1}{3}B_{+}-B_{-}.
\end{equation}
In terms of the above new variables, the matrices of the principal part of the system take the form
\begin{equation*}
                    \begin{pmatrix}
			    2\bmath{e}_{0} & 0 & 0 & 0 & -D_{1} & D_{2} & 0 & 0 & -D_{3} & D_{3} \\
			    0 & 2\bmath{e}_{0} & 0 & D_{1} & 0 & -D_{3} & 0 & 0 & D_{2} & -D_{2} \\
		            0 & 0 & 2\bmath{e}_{0} & -D_{2} & D_{3} & 0 & 0 & 0 & 0 & 2D_{1} \\
                            0 & D_{1} & -D_{2} & 2\bmath{e}_{0} & 0 & 0 & D_{3} & D_{3} & 0 & 0 \\
                            -D_{1} & 0 & D_{3} & 0 & 2\bmath{e}_{0} & 0 & -D_{2} & D_{2} & 0 & 0 \\
                            D_{2} & -D_{3} & 0 & 0 & 0 & 2\bmath{e}_{0} & 0 & -2D_{1} & 0 & 0 \\
       		            0 & 0 & 0 & D_{3} & -D_{2} & 0 & \bmath{e}_{0} & 0 & 0 & 0 \\
		            0 & 0 & 0 & D_{3} & D_{2} & -2D_{1} & 0 & \bmath{e}_{0} & 0 & 0 \\
         		    -D_{3} & D_{2} & 0 & 0 & 0 & 0 & 0 & 0 & \bmath{e}_{0} & 0 \\
                            D_{3} & -D_{2} & 2D_{1} & 0 & 0 & 0 & 0 & 0 & 0 & \bmath{e}_{0} \\
		    \end{pmatrix}
                    \begin{pmatrix}
			    E_{12} \\
                            E_{13} \\
                            E_{23} \\
                            B_{12} \\
                            B_{13} \\
                            B_{23} \\
                            E_{+} \\
                            E_{-} \\
                            B_{+} \\
                            B_{-}
		    \end{pmatrix}
\end{equation*}
and
\begin{equation*}
 \begin{pmatrix}
  \bmath{e}_{0} & 0 & 0 & -D_{2} & -D_{3} & 0 & 2D_{1} & 0 & -D_{1}  \\
  0 & \bmath{e}_{0} & 0 & -D_{1} & 0 & -D_{3} & -D_2 & -3D_{2} & -D_{2} \\
  0 & 0 & \bmath{e}_{0} & 0 & -D_{1} & -D_{2} & -D_3 &  3D_3 & -D_{3}\\
  -D_{2} & -D_{1} & 0 & 2\bmath{e}_{0} & 0 & 0 & 0 & 0  & 0 \\
  -D_{3} & 0 & -D_{1} & 0 & 2\bmath{e}_{0} & 0 & 0 & 0  & 0  \\
  0 & -D_{3} & -D_{2} & 0 & 0 & 2\bmath{e}_{0} & 0 & 0  & 0 \\
  2D_{1} & -D_{2} & -D_{3} & 0 & 0 & 0 & 6\bmath{e}_{0} & 0 & 0 \\
  0 & -3D_{2} & 3D_{3} & 0 & 0 & 0 & 0 & 18\bmath{e}_{0} & 0 \\
 -D_{1} & -D_{2} & -D_{3} & 0 & 0 & 0 & 0 & 0 & 3\bmath{e}_{0}
\end{pmatrix}
 \begin{pmatrix}
  a_{1} \\
  a_{2} \\
  a_{3} \\
  \left(\chi^{ST}\right)_{12} \\
  \left(\chi^{ST}\right)_{13} \\
  \left(\chi^{ST}\right)_{23} \\
  \chi_{+} \\
  \chi_{-} \\ 
  H
 \end{pmatrix},
\end{equation*}
which are symmetric. 
\subsection{The background solution}
As is well known, the metric of a FLRW spacetime can be written as 
\begin{equation*}
 \bmath{g}_{\text{FLRW}}=-\mbox{d}t^{2}+\left(\frac{a(t)}{\omega}\right)^{2}\delta_{ij}
 \mbox{d}x^{i}\mbox{d}x^{j},
\end{equation*}
where $a(t)$ is the scale factor and
\begin{equation*}
 \omega=1+\frac{k}{4}\delta_{ij}x^{i}x^{j}, \quad \quad\partial_{i}\omega=\frac{k}{2}x_{i},
\end{equation*}
with the constant $k=-1,0,1$ being the curvature of the spatial hypersurfaces. 
Since the metric is conformally flat, it follows that 
\[
\mathring{E}_{bd}=\mathring{B}_{bd}=0.
\]
Now, the gauge conditions for the frame are satisfied if one sets 
\begin{equation*}
 \mathring{e}_{0}{}^{\mu}=\delta_{0}{}^\mu,\qquad\mathring{e}_{b}{}^\mu={\left(\frac{\omega}{a}\right)}\delta_{b}{}^\mu,\qquad b=1,2,3
\end{equation*}
so that the spatial connection coefficients are given by
\begin{equation*}
 \mathring{\gamma}^{c}{}_{bd}=\frac{k}{2a^{2}}\left(h_{db}x^{c}-h_{d}{}^{c}x_{b}\right),\quad b,c,d=1,2,3
\end{equation*}
with $x^{\mu}=(\omega/a)\delta^{\mu}{}_{c}x^{c}$. The remaining non-vanishing connection coefficients are
\begin{equation*}
 \mathring{\gamma}^{0}{}_{bd}=\mathring{\chi}_{db}=\mathring{H}h_{bd}, \quad  \quad \mathring{\gamma}^{b}{}_{0d}=\mathring{\chi}_{d}{}^{b}=\mathring{H}h_{d}{}^{b},\quad b,d=1,2,3
\end{equation*}
and, using \eqref{hubble} and \eqref{Decomposition_2nFF}, we write
\begin{equation*}
\mathring{\chi}=3\mathring{H},~~~~~~\mathring\chi_{[bd]}=a_{b}=0,\quad\text{and}\quad \mathring\chi_{(bd)}=0\quad \text{for} \quad b\neq d,
\end{equation*}
where, in this case, $\displaystyle{\mathring{H}(t)=\frac{1}{a}\frac{da}{dt}}$.
The Einstein-scalar field system thus reduces to the evolution equations 
\begin{equation}
\begin{aligned}
 \frac{\mbox{d}\mathring{\phi}}{\mbox{d}t}&=\mathring{\psi}, \\
 \frac{\mbox{d}\mathring{\psi}}{\mbox{d}t}&=-3\mathring{H}\mathring{\psi}-\frac{\mbox{d}\mathcal{\mathring V}}{\mbox{d}\mathring{\phi}}, \\
 \frac{\mbox{d}\mathring{H}}{\mbox{d}t}&=-\mathring{H}^{2}-\frac{1}{3}\mathring{\psi}^{2}+\frac{1}{3}\mathcal{\mathring V}(\mathring{\phi}),
\end{aligned}
\label{evol_background}
\end{equation}
subject to the Friedman-scalar field constraint equation
\begin{equation}
 \mathring{H}^{2}=\frac{1}{3}\left[\frac{1}{2}\mathring{\psi}^{2}+\mathcal{\mathring V}(\mathring{\phi})\right]-\frac{k}{{a}^{2}} .
\label{Friedmannequation}
\end{equation}
\subsection{Linearised evolution equations}
\label{LinearSection}
In this subsection we derive the linearised system associated to the
nonlinear equations of Theorem \ref{Theorem:wellposedness}, for the
case of a FLRW background with a self-interacting scalar field. In order
to perform the linearisation procedure we compute
\begin{equation}
\left.\frac{\mbox{d}\mathbf{u}^{\epsilon}}{\mbox{d}\epsilon}\right |_{\epsilon=0} \nonumber
\end{equation} 
and drop all (nonlinear) terms of coupled perturbations. In this way, we obtain the following linearised system for $b\leq d$
\begin{eqnarray*}
&&\partial_{t}\breve{\phi}=\breve{\psi}, \\
&& \partial_{t}\breve{\psi}=-\left({\frac{\mbox{d}^{2}\mathcal{\mathring{V}}}{\mbox{d}\mathring{\phi}^{2}}}\right)\breve{\phi}-3\mathring{H}\breve{\psi}-3\mathring{\psi}\breve{H},\\
&& 3\partial_{t}\breve{H}-\left(\frac{\omega}{{a}}\right)\delta_{p}{}^{j}\partial_{j}\breve{a}^{p}=\frac{\mbox{d}\mathcal{V}}{\mbox{d}\mathring{\phi}}\breve{\phi}-2\mathring{\psi}\breve{\psi}-6\mathring{H}\breve{H}-\frac{k}{a^{2}}x_{p}\breve{a}^{p}, \\
&& 2\partial_{t}\left(\breve{\chi}^{ST}\right)_{bd}-2\left(\frac{\omega}{{a}}\right)\delta_{(b}{}^{j}\partial_{j}\breve{a}_{d)}+2\left(\frac{\omega}{{a}}\right)\frac{h_{bd}}{3}\delta_{p}{}^{j}\partial_{j}\breve{a}^{p}=-4\mathring{H}\left(\breve{\chi}^{ST}\right)_{bd}-2\breve{E}_{bd}+\frac{k}{a^{2}}x_{(b}\breve{a}_{d)}-\frac{h_{bd}}{3}\frac{k}{a^{2}}x_{p}\breve{a}^{p}, \\
&& \partial_{t}\breve{a}_{c}-\left(\frac{\omega}{a}\right)\delta^{\,\,\,j}_{p}\partial_{j}\left(\breve{\chi}^{ST}\right)_{c}{}^{p}-\left(\frac{\omega}{a}\right)\delta^{\,\,\,j}_{c}\partial_{j}\breve{H}=2{\left(\mathring{H}+\frac{\mathring{\mathcal{V}}^{\prime}}{\mathring{\psi}}\right)}\breve{a}_{c}+\frac{d\mathring{H}}{dt}\breve{\bar{e}}^{\,\,\,0}_{c}-\frac{3}{2}\frac{k}{a^{2}}x_{f}\left(\breve{\chi}^{ST}\right)_{c}{}^{f}, \\
&& \partial_{t}\breve{\bar{e}}_{c}^{\,\,\mu}=-\mathring{H}\breve{\bar{e}}_{c}^{\,\,\mu}-\left(\frac{\omega}{a}\right)\delta_{c}^{\,\,\mu}\breve{H}-\left(\frac{\omega}{a}\right)\delta_{b}^{\,\,\mu}\left(\breve{\chi}^{ST}\right)_{c}{}^{b}+\delta_{0}^{\,\,\mu}\breve{a}_{c}, \\
&&  2\partial_{t}\breve{E}_{bd}-2\left(\frac{\omega}{a}\right)\epsilon^{pa}{}_{(b|}\delta_{a}{}^{j}\partial_{j}\breve{B}_{p|d)}=-2\mathring{H}\breve{E}_{bd}-\mathring{\psi}^{2}\left(\breve{\chi}^{ST}\right)_{bd}+\frac{k}{a^{2}}\left(x_{p}\breve{B}_{a(b}\epsilon_{d)}^{\,\,\,pa}+\epsilon_{\,\,\,(b}^{pa}x_{d)}\breve{B}_{pa}\right),\\
&& 2\partial_{t}\breve{B}_{bd}-2\left(\frac{\omega}{a}\right)\epsilon^{ap}{}_{(b|}\delta_{a}{}^{j}\partial_{j}\breve{E}_{p|d)}=-2\mathring{H}\breve{B}_{bd}+\frac{k}{a^{2}}\left(x_{p}\breve{E}_{a(b}\epsilon_{d)}^{\,\,\,ap}+\epsilon_{\,\,\,(b}^{ap}x_{d)}\breve{E}_{pa}\right), \\
&& \partial_{t}\breve{\bar{\gamma}}^{c}_{\,\,bd}=-\mathring{H}\breve{\bar{\gamma}}^{c}_{\,\,bd}-\frac{k}{2a^{2}}\left(h_{db}x^{c}-\delta^{c}_{d}x_{b}\right)\breve{H}-\frac{k}{2a^{2}}\left[x^{c}\left(\breve{\chi}^{ST}\right)_{db}-x_{b}\left(\breve{\chi}^{ST}\right)^{\,\,c}_{d}\right], \\
&& \hspace{2cm}-\mathring{H}\left(h_{db}\breve{a}^{c}-\delta^{c}_{d}\breve{a}_{b}\right)+\epsilon^{pc}_{\,\,\,\,b}\breve{B}_{dp}.   
\end{eqnarray*}
For a flat ($k=0$) background, using the variables \eqref{NEWVAR}, the linearised system has the form
\begin{equation}
\label{linearised-system}
 \mathring{\mathbf{A}}^{0}\partial_{t}\breve{\mathbf{u}}-a^{-1}(t)\mathring{\mathbf{A}}^{j}\partial_{j}\breve{\mathbf{u}}=\mathring{\mathbf{B}}(t)\breve{\mathbf{u}}
\end{equation}
where $\mathring{\mathbf{A}}^{0}\partial_{t}\breve{\mathbf{u}}-a^{-1}(t)\mathring{\mathbf{A}}^{j}\partial_{j}\breve{\mathbf{u}}$ is given by 
\begin{equation*}
                    \begin{pmatrix}
			    2\partial_{t} & 0 & 0 & 0 & -a^{-1}\partial_{1} & a^{-1}\partial_{2} & 0 & 0 & -a^{-1}\partial_{3} & -a^{-1}\partial_{3} \\
			    0 & 2\partial_{t} & 0 & a^{-1}\partial_{1} & 0 & -a^{-1}\partial_{3} & 0 & 0 & a^{-1}\partial_{2} & -a^{-1}\partial_{2} \\
		            0 & 0 & 2\partial_{t} & -a^{-1}\partial_{2} & a^{-1}\partial_{3} & 0 & 0 & 0 & 0 & 2a^{-1}\partial_{1} \\
                            0 & a^{-1}\partial_{1} & -a^{-1}\partial_{2} & 2\partial_{t} & 0 & 0 & a^{-1}\partial_{3} & a^{-1}\partial_{3} & 0 & 0 \\
                            -a^{-1}\partial_{1} & 0 & a^{-1}\partial_{3} & 0 & 2\partial_{t} & 0 & -a^{-1}\partial_{2} & a^{-1}\partial_{2} & 0 & 0 \\
                            a^{-1}\partial_{2} & -a^{-1}\partial_{3} & 0 & 0 & 0 & 2\partial_{t} & 0 & -2a^{-1}\partial_{1} & 0 & 0 \\
       		            0 & 0 & 0 & a^{-1}\partial_{3} & -a^{-1}\partial_{2} & 0 & \frac{2}{3}\partial_{t} & 0 & 0 & 0 \\
		            0 & 0 & 0 & a^{-1}\partial_{3} & a^{-1}\partial_{2} & -2a^{-1}\partial_{1} & 0 & 2\partial_{t} & 0 & 0 \\
         		    -a^{-1}\partial_{3} & a^{-1}\partial_{2} & 0 & 0 & 0 & 0 & 0 & 0 & \frac{2}{3}\partial_{t} & 0 \\
                            -a^{-1}\partial_{3} & -a^{-1}\partial_{2} & 2a^{-1}\partial_{1} & 0 & 0 & 0 & 0 & 0 & 0 & 2\partial_{t} \\
		    \end{pmatrix}
                    \begin{pmatrix}
			    \breve{E}_{12} \\
                            \breve{E}_{13} \\
                            \breve{E}_{23} \\
                            \breve{B}_{12} \\
                            \breve{B}_{13} \\
                            \breve{B}_{23} \\
                            \breve{E}_{+} \\
                            \breve{E}_{-} \\
                            \breve{B}_{+} \\
                            \breve{B}_{-}
		    \end{pmatrix},
\end{equation*}
and
\begin{equation*}
 \begin{pmatrix}
  3\partial_{t} & 0 & 0 & 0 & -a^{-1}\partial_{1} & -a^{-1}\partial_{2} & -a^{-1}\partial_{3} & 0 & 0  \\
  0 & 2\partial_{t} & 0 & 0 & -a^{-1}\partial_{2} & -a^{-1}\partial_{1} & 0 & 0 & 0 \\
  0 & 0 & 2\partial_{t} & 0 & -a^{-1}\partial_{3} & 0 & -a^{-1}\partial_{1} & 0 & 0 \\
  0 & 0 & 0 & 2\partial_{t} & 0 & -a^{-1}\partial_{3} & -a^{-1}\partial_{2} & 0 & 0 \\
  -a^{-1}\partial_{1} & -a^{-1}\partial_{2} & -a^{-1}\partial_{3} & 0 & \partial_{t} & 0 & 0  & 2\partial_{1} & 0 &   \\
  -a^{-1}\partial_{2} & -a^{-1}\partial_{1} & 0 & -a^{-1}\partial_{3} & 0 & \partial_{t} & 0 & -\partial_{2} & -3\partial_{2} &  \\
  -a^{-1}\partial_{3} & 0 & -a^{-1}\partial_{1} & -a^{-1}\partial_{2} & 0 & 0 & \partial_{t} & -\partial_{3} &  3\partial_{3} & \\
  0 & 0 & 0 & 0 & 2a^{-1}\partial_{1} & -a^{-1}\partial_{2} & -\partial_{3}  & 6\partial_{t} & 0 \\
  0 & 0 & 0 & 0 & 0 & -3a^{-1}\partial_{2} & 3a^{-1}\partial_{3} & 0 & 18\partial_{t} \\

\end{pmatrix}
 \begin{pmatrix}
  \breve{H} \\
  \left(\breve{\chi}^{ST}\right)_{12} \\
  \left(\breve{\chi}^{ST}\right)_{13} \\
  \left(\breve{\chi}^{ST}\right)_{23} \\
  \breve{a}_{1} \\
  \breve{a}_{2} \\
  \breve{a}_{3} \\
  \breve{\chi}_{+} \\
  \breve{\chi}_{-} \\ 
 \end{pmatrix}.
\end{equation*}
If
\begin{equation*}
 \breve{\mathbf{u}}^{T}=\left[\breve{\phi}\,\,\breve{\psi}\,\,\breve{H}\,\,(\breve{\chi}^{ST})_{bd}\,\,\breve{E}_{bd}\,\,\breve{\chi}_{+}\,\,\breve{E}_{+}\,\,\breve{\chi}_{-}\,\,\breve{E}_{-}\,\,\,\breve{a}_{c}\,\,\breve{e}_{c}^{\,\,0}\,\,\breve{B}_{bd}\,\,\breve{B}_{+}\,\,\breve{B}_{-}\,\,\breve{e}_{c}^{\,\,j}\,\,\breve{\gamma}^{c}_{\,\,bd}\right]
\end{equation*}
then the matrix $\mathring{\mathbf{B}}(t)$ is explicitly given by
\begin{equation*}
 \left(\begin{smallmatrix}
 	\mathbf{B}^{(1)}_{3\times3} & 0 & 0 & 0 & 0 & 0 & 0 & 0 & 0 & 0 & 0 & 0 & 0 & 0\\
 	0 & -4\mathring{H}\mathbf{I}_{3\times3} & -2\mathbf{I}_{3\times3} & 0 & 0 & 0 & 0 & 0 & 0 & 0 & 0 & 0 & 0 & 0\\
   	0 & -\mathring{\psi}^{2}\mathbf{I}_{3\times3} & -2\mathring H\mathbf{I}_{3\times3} & 0 & 0 & 0 & 0 & 0 & 0 & 0 & 0 & 0 & 0 & 0\\
 	0 & 0 & 0 & -12\mathring{H} & -2 & 0 & 0 & 0 & 0 & 0 & 0 & 0 & 0 & 0 \\
 	0 & 0 & 0 & -\mathring{\psi}^{2} & -\frac{2}{3}\mathring{H} & 0 & 0 & 0 & 0 & 0 & 0 & 0 & 0 & 0\\
 	0 & 0 & 0 & 0 & 0 & -36\mathring{H} & -6 & 0 & 0 & 0 & 0 & 0 & 0 & 0 \\
 	0 & 0 & 0 & 0 & 0 & -3\mathring{\psi}^{2} & -2\mathring{H} & 0 & 0 & 0 & 0 & 0 & 0 & 0 \\
 	0 & 0 & 0 & 0 & 0 & 0 & 0 & 2\left({\frac{\mathring{\mathcal{V}}^{\prime}}{\mathring{\psi}}}+\mathring{H}\right)\mathbf{I}_{3\times3} & \frac{d\mathring{H}}{dt}\mathbf{I}_{3\times3} & 0 & 0 & 0 & 0 & 0 \\
        0 & 0 & 0 & 0 & 0 & 0 & 0 & \mathbf{I}_{3\times3} & -\mathring{H}\mathbf{I}_{3\times3} & 0 & 0 & 0 & 0 & 0 \\
        0 & 0 & 0 & 0 & 0 & 0 & 0 & 0 & 0 & -2\mathring{H}\mathbf{I}_{3\times3} & 0 & 0 & 0 & 0 \\
        0 & 0 & 0 & 0 & 0 & 0 & 0 & 0 & 0 & 0 & -\frac{2}{3}\mathring{H} & 0 & 0 & 0  \\
        0 & 0 & 0 & 0 & 0 & 0 & 0 & 0 & 0 & 0 & 0 & -2\mathring{H} & 0 & 0 \\
        \mathbf{B}^{(2)}_{3\times3} & \mathbf{B}^{(3)}_{3\times3} & 0 & \mathbf{B}^{(4)}_{3\times1} & 0 & \mathbf{B}^{(5)}_{3\times1} & 0 & 0 & 0 & 0 & 0 & 0 & -\mathring{H}\mathbf{I}_{3\times3} & 0  \\
        0 & 0 & 0 & 0 & 0 & 0 & 0 & -\mathring{H}\mathbf{B}^{(6)}_{9\times3} & 0 & \mathbf{B}^{(7)}_{9\times3} & \mathbf{B}^{(8)}_{9\times1} & \mathbf{B}^{(9)}_{9\times1} & 0 & -\mathring{H}\mathbf{I}_{9\times9}  \\
        \end{smallmatrix}\right)
\end{equation*}
 where
 \begin{equation*}
 \mathbf{B}^{(1)}_{3\times3}=\begin{pmatrix}
                     0 & 1 & 0 \\
		     -\mathring{\mathcal{V}}^{\prime\prime} & -3\mathring{H} & -3\mathring{\psi} \\
                     \mathring{\mathcal{V}}^{\prime} & -2\mathring{\psi} & -6\mathring{H}
                    \end{pmatrix},\quad 
  \mathbf{B}^{(2)}_{3\times3}=\begin{pmatrix}
                     0 & 0 & -a^{-1}\delta^{j}_{1} \\
		     0 & 0 & -a^{-1}\delta^{j}_{2} \\
                     0 & 0 & -a^{-1}\delta^{j}_{3}
		\end{pmatrix}, \quad 
  \mathbf{B}^{(3)}_{3\times3}=\begin{pmatrix}
                     -a^{-1}\delta^{j}_{2} & -a^{-1}\delta^{j}_{3} & 0 \\
		     -a^{-1}\delta^{j}_{1} & 0 & -a^{-1}\delta^{j}_{3} \\
                     0 & -a^{-1}\delta^{j}_{1} & -a^{-1}\delta^{j}_{2} 
		\end{pmatrix},   
\end{equation*}
\begin{equation*}
\mathbf{B}^{(4)}_{3\times1}=\begin{pmatrix}
                       2a^{-1}\delta^{j}_{1} \\
		       -a^{-1}\delta^{j}_{2} \\
		       -a^{-1}\delta^{j}_{3}		      
                     \end{pmatrix},\quad		 
\mathbf{B}^{(5)}_{3\times1}=\begin{pmatrix}
                       0 \\
		       -3a^{-1}\delta^{j}_{2} \\
		       3a^{-1}\delta^{j}_{3}		      
                     \end{pmatrix},\quad		                     
\end{equation*}
\begin{equation*}
\mathbf{B}^{(6)}_{9\times3}=\begin{pmatrix}
                     1 & 0 & 0 \\
		     1 & 0 & 0 \\
		     0 & 0 & 0 \\
                     0 & 1 & 0 \\
		     0 & 1 & 0 \\
		     0 & 0 & 0 \\
		     0 & 0 & 1 \\
		     0 & 0 & 1 \\
                     0 & 0 & 0
		    \end{pmatrix},\quad 
\mathbf{B}^{(7)}_{9\times3}=\begin{pmatrix}
                     0 & 0 & 1 \\
		     0 & 0 & -1 \\
		     0 & 0 & 0 \\
		     0 & -1 & 0 \\
		     0 & 1 & 0 \\
		     0 & 0 & 0 \\
                     1 & 0 & 0 \\
		     -1 & 0 & 0 \\
		     0 & 0 & 0
                    \end{pmatrix},\quad
\mathbf{B}^{(8)}_{9\times1}=\begin{pmatrix}
                     0  \\
		     0  \\
		     -1  \\
		     0  \\
		     0  \\
		     0  \\
                     0  \\
		     0  \\
		     1 
                    \end{pmatrix},\quad
\mathbf{B}^{(9)}_{9\times1}=\begin{pmatrix}
                     0  \\
		     0  \\
		     \frac{1}{3}  \\
		     0  \\
		     0  \\
		     -\frac{2}{3}  \\
                     0  \\
		     0  \\
		     \frac{1}{3} 
                    \end{pmatrix}.
\end{equation*} 
\subsection{Asymptotic exponential decay of nonlinear perturbations}

In this section, we show how well-known results from the theory of nonlinear symmetric hyperbolic systems can be generalised 
and applied to the analysis of the asymptotic exponential decay of nonlinear perturbations of the flat FLRW reference solution 
to the Einstein-nonlinear scalar field system. From the analysis carried out in the previous sections, 
we consider the initial-value problem for the nonlinear perturbations of the form
\begin{equation}
 \begin{aligned}
 &\left(\mathring{\mathbf{A}}^{0}-\epsilon \breve{{\bf A}}^{0}(\mathring{{\bf u}},\breve{{\bf u}},\epsilon)\right)\partial_{t}\breve{{\bf u}}-\left(\mathring{\mathbf{A}}^{j}(\mathring{{\bf u}})+\epsilon \breve{{\bf A}}^{j}(\mathring{{\bf u}},\breve{{\bf u}},\epsilon)\right)\partial_{j}\breve{{\bf u}}=\left(\mathring{{\bf B}}(\mathring{{\bf u}})+\epsilon \breve{{\bf B}}(\mathring{{\bf u}},\breve{{\bf u}},\epsilon)\right)\breve{{\bf u}}, \\
 &\breve{{\bf u}}(\bmath{x},0)=\breve{\mathbf{u}}_{0}(\bmath{x}),
\end{aligned}
\label{IVP_nonlinear_pert}
\end{equation}
where $\mathring{{\bf u}}=\mathring{{\bf u}}(t)$.
The matrix $\mathring{\mathbf{A}}^{0}$ is symmetric with positive entries, 
and $\breve{\mathbf{A}}^{0}$, $\mathring{\mathbf{A}}^{j}$ and $\breve{\mathbf{A}}^{j}$, $j=1,2,3$, are symmetric.

The nonlinear stability of the solutions to the Cauchy problem
\eqref{IVP_nonlinear_pert}, when the coefficients of the linearized
system ($\epsilon=0$) are constant matrices, has been studied
extensively in \cite{KREISSBOOK,KS97,KL98,KOR98,Ort01}. The
key ingredient in the analysis is the requirement that the eigenvalues
of the matrix $\mathring{\mathbf{B}}$ have a negative real part. 
It is the purpose of the next subsection to show how this \emph{stability eigenvalue condition} can be 
generalised under certain assumptions for problems where the matrices of the linearised system have entries which 
are smooth functions of time $t$. This will be the key result of next section and is given in Lemma \ref{AsymptEigenValue}. 
This lemma allows us to write a stability theorem, given by Theorem~\ref{StabilityTheorem}, whose proof follows closely the 
methods of \cite{KL98,KOR98} which, in turn, are based on \cite{KREISSBOOK,KS97}. 
We shall then omit details in some parts of the proof but shall give references where the missing steps can be found.
The last subsection contains our main theorem, where the stability Theorem~\ref{StabilityTheorem} is applied to the 
present case. The results are summarised in Theorem~\ref{Theorem:Main}.

\subsubsection{A stability theorem}
In what follows, we denote by $\langle\cdot,\cdot\rangle$ the inner product in
$L^{2}\left(\mathbb{T}^{n}\right)$
\begin{equation*}
 \langle \mathbf{f},\mathbf{g}\rangle \equiv \int_{\mathbb{T}^{n}}\mathbf{f}^{T}\mathbf{g}\mbox{d}\bmath{x},
\end{equation*}
where ${\bf f, g}:\mathbb{T}^{n}\rightarrow\mathbb{R}^{s}$, and the corresponding norm
\begin{equation*}
\|{\bf f}\|^{2}_{L^{2}\left(\mathbb{T}^{n}\right)} \equiv \langle \mathbf{f},\mathbf{f}\rangle.
\end{equation*}
We use the notation
\[
\partial^{\alpha}_{x} \mathbf f=\frac{\partial^{|\alpha|}\mathbf f}{\partial
(x_{1})^{\alpha_{1}}...\partial (x_{n})^{\alpha_{n}}},
\]
where $\alpha=(\alpha_1,..,\alpha_n )$ is a multi-index with respect to $ x =(x_1, ..,x_n)$, for non-negative integers $\alpha_i$. 
Let $H^{k}\left(\mathbb{T}^{n};\mathbb{R}^{s}\right)$ be the
space of all summable functions $\mathbf f$ such that, for
each multi-index $|\alpha|\leq k$, $\partial_x^{\alpha}{\bf f}$
exists in the weak sense and belongs to
$L^{2}\left(\mathbb{T}^{n}\right)$. The norm in $H^{k}\left(\mathbb{T}^{n};\mathbb{R}^{s}\right)$ is
\begin{equation*}
\|{\bf f}\|^2_{H^{k}(\mathbb{T}^{n})} \equiv \sum_{|\alpha|=0}^{k}\int_{\mathbb{T}^{n}}(\partial_x^{\alpha}{\bf f})^{2}\mbox{d}\bmath{x},
\end{equation*}
We note that even when $\mathbf f\in C^\infty (\mathbb T^n\times I)$ for some time interval $I$, it should be understood 
that $\partial_x^{\alpha}{\bf f}$ means differentiation with respect to the spatial variables only.

We shall now recall usual assumptions for short time existence theorems in this context (see e.g. \cite{Ort01}):
\begin{Assumption}
\label{ASS1}
If
\begin{equation*}
 |\mathring{\mathbf{u}}(t)|\leq c 
\end{equation*}
\begin{equation*}
 |\breve{\mathbf{u}}(\bmath{x},t)|\leq c 
\end{equation*}
for some $c>0$, then for every $k=0,1,2,...$, there are constants $p_{A^ 0,k}$, $p_{A,k}$, $p_{B,k}$ and $K(c,k)$ 
such that\footnote{Here $\partial^{|\beta|}_{u}=\frac{\partial^{|\beta|}}{\partial u^{\beta_1}_{1}...\partial u^{\beta_s}_{s}}$ 
denote multi-indices with respect to $\mathbf{u}$.} 
 \begin{equation*}
 \begin{split}
|\breve{\mathbf{A}}^{0}(\mathring{{\bf u}},\breve{{\bf u}},\epsilon)|\leq p_{A^{0},k}|\breve{\mathbf{u}}(\bmath{x},t)|,\qquad |\breve{\mathbf{A}}^{j}(\mathring{{\bf u}},\breve{{\bf u}},\epsilon)|\leq p_{A,k}|\breve{\mathbf{u}}(\bmath{x},t)|,  &\qquad|\breve{\mathbf{B}}(\mathring{{\bf u}},\breve{{\bf u}},\epsilon)|\leq p_{B,k}|\breve{\mathbf{u}}(\bmath{x},t)|, \\
 \end{split}
\end{equation*}
\begin{equation*}
 \begin{split}
     |\partial^{\alpha}_{x}\partial^{\beta}_{u}\breve{\mathbf{A}}^{0}|\leq K(c,k),\qquad     \sum^{n=3}_{j=1} |\partial^{\alpha}_{x}\partial^{\beta}_{u}\breve{\mathbf{A}}^{j}|\leq K(c,k) ,\qquad  |\partial^{\alpha}_{x}\partial^{\beta}_{u}\breve{\mathbf{B}}|\leq K(c,k),   
 \end{split}
\end{equation*}
for all multi-indices $\alpha$ and $\beta$ with $|\alpha|+|\beta|=k$.
For the initial data $\breve{\mathbf{u}}_{0}$, the corresponding estimates hold.
\label{AA1}
\end{Assumption} 
Now, let ${\bf \breve u}:\mathbb{T}^{n}\times I\rightarrow\mathbb{R}^{s}$ be the unknown of \eqref{IVP_nonlinear_pert}.
The subsequent argument makes frequent use of the usual Sobolev inequalities (see e.g. \cite{KREISSBOOK})
\begin{equation}
\label{sobo1}
  \|\breve{\mathbf{u}}\|_{L^{\infty}(\mathbb{T}^{n})}\leq C_{k,n} \|\breve{\mathbf{u}}\|_{H^{k}(\mathbb{T}^{n})}\quad\text{if}\quad k>\frac{n}{2}
\end{equation}
and
\begin{equation}
\label{sobo2}
 \|\partial_x^{\alpha}\breve{\mathbf{u}}\|_{L^{\infty}(\mathbb{T}^{n})}\leq C_{k,n} \|\breve{\mathbf{u}}\|_{H^{k}(\mathbb{T}^{n})}\quad\text{if}\quad k\geq |\alpha|+[n/2]+1
\end{equation}
where $[n/2]$ denotes the largest integer not greater than $n/2$ and
$C_{k,n}$ are some positive constants. 
Under Assumption \ref{AA1}, and using the above Sobolev inequalities, one has the
following estimates based on the chain rule 
\begin{equation}
\label{chain}
 \|\partial^{\alpha}_{x}\breve{\mathbf{A}}^{j}\|_{L^{\infty}(\mathbb{T}^{n})}\leq C\, K(c,k)(1+m^{k-1})\|\breve{\mathbf{u}}\|_{H^{k}(\mathbb{T}^{n})}
\end{equation}
where
$m=\max_{\alpha}\left\{\|\partial_x^{\alpha}\breve{\mathbf{u}}\|_{L^{\infty}(\mathbb{T}^{n})}\;:\;k>|\alpha|+[\frac{n}{2}]+1\right\}$
and the constant $C$ is independent of $\breve{\mathbf{A}}^{j}$ and
$\breve{\mathbf{u}}$, see \cite{KREISSBOOK}.

In what follows we introduce an extra assumption, followed by a generalisation of the \emph{stability eigenvalue condition} \cite{KL98}, 
for the system \eqref{IVP_nonlinear_pert}, in which the linearised matrices are not constant but depend on time. 
These will then be used to derive Lemma \ref{AsymptEigenValue} which, in turn, is crucial for the construction of our energy estimates. 
The lemma, which generalises techniques of the proofs of Lemma 2.1 and Theorem 2.2 of \cite{KL98}, states that, 
under certain conditions, the eigenvalues can be estimated by their values at infinity. 
\begin{Assumption}
\label{ASS2}
 The coefficients $\mathring{\mathbf{A}}^{j}\left(\mathring{\mathbf{u}}(t)\right)$ and $\mathring{\mathbf{B}}\left(\mathring{\mathbf{u}}(t)\right)$ 
 are bounded 
 and converge to a finite limit $\mathring{\mathbf{A}}^{j}_{\infty}$ and $\mathring{\mathbf{B}}_{\infty}$, as $t\rightarrow+\infty$.
\end{Assumption}
\begin{Assumption}
There exists a constant $\delta_\infty>0$ such that all the eigenvalues $\lambda_\infty$ of $\mathring{{\bf B}}_\infty$ satisfy 
\begin{equation}
 \text{Re}\left(\lambda_\infty\right)\leq-\delta_\infty. 
\end{equation}
\label{ASS3}
\end{Assumption}
\begin{lem}
\label{AsymptEigenValue}
If Assumption~\ref{ASS2} and Assumption \ref{ASS3} hold, then there exists a
time $T>0$ and a constant $\delta_1>0$ 
such that, for all $t\in[T,\infty)$,
\begin{equation}
 \mathbf{S}_{\infty}\mathring{{\bf B}}(t)+\mathring{{\bf B}}^{T}(t)\mathbf{S}_{\infty}\leq -2\delta_{1}\mathbf{S}_{\infty},
 \end{equation}
 with $\mathbf{S}_{\infty}$ being a positive definite Hermitian matrix with constant entries.
 \end{lem}
\begin{proof}
If Assumption~\ref{ASS2} holds, then there is a function $\delta(t)$ such that $\delta_{\infty}\equiv \lim_{t\to  \infty} \delta(t)$. 
Futhermore, if Assumption \ref{ASS3} holds, then there exists a time $T>0$ and a constant $\delta>0$ such that for 
all $t\in[T,\infty)$
\begin{equation}
\label{SEC_infty}
 \text{Re}\left(\lambda(t)\right)=\text{Re}\left(\lambda_{\infty}+\Delta(t)\right)\leq -\delta_{\infty}+\text{Re}\left(\Delta(t)\right) \leq -\delta,
\end{equation}
where $\Delta(t)=\lambda(t)-\lambda_{\infty}$. Thus, if the data is
given at $t_{0}=T$, the eigenvalues can be estimated by their values
at infinity. 
If $\mathring{{\bf B}}(t)$ is diagonal,
then this condition can be directly translated into the following
crucial inequality for the energy estimates (see also Lemma 2.1 of \cite{KL98})
\begin{equation*}
 \mathring{{\bf B}}(t)+\mathring{{\bf B}}^{T}(t)\leq2 \text{Re}\left(\lambda(t)\right)\mathbf{I_{d}}\leq-2\delta\mathbf{I_{d}}.
\end{equation*}
Otherwise, there exists a positive definite Hermitian matrix 
$$
\mathbf{S}_{\infty}=\mathbf{Q}^{\dag}_{\infty}\mathbf{Q}_{\infty}=\left(\left(\mathbf{U}_{\infty}\mathbf{D}\right)^{-1}\right)^{\dag}\left(\mathbf{U}_{\infty}\mathbf{D}\right)^{-1}
$$
where $\mathbf{U}_{\infty}$ is an unitary matrix which puts
$\mathring{\mathbf{B}}_{\infty}$ in its Schur's form:
$$
\mathbf{U}_{\infty}\mathring{\mathbf{B}}_{\infty}\mathbf{U}^{-1}_{\infty}=\mathbf{\Lambda}_{\infty}+\mathbf{R}_{\infty}
$$
with $\mathbf{\Lambda}_{\infty}$ being a diagonal matrix whose entries are the
eigenvalues $\lambda_{\infty}$ and $\mathbf{R}_{\infty}$ is an upper
triangular matrix. In turn, the matrix $\mathbf{D}$ is a diagonal matrix with
arbitrary positive constant entries such that
$$
\mathbf{Q}_{\infty}\mathring{\mathbf{B}}_{\infty}\mathbf{Q}^{-1}_{\infty}=\mathbf{D}^{-1}\mathbf{U}_{\infty}\mathring{\mathbf{B}}_{\infty}\mathbf{U}^{-1}_{\infty}\mathbf{D}=\mathbf{\Lambda}_{\infty}+\mathbf{D}^{-1}\mathbf{R}_{\infty}\mathbf{D}.
$$
Then, we have
\begin{equation*}
\begin{aligned}
 \mathbf{S}_{\infty}\mathring{{\bf B}}(t)+\mathring{{\bf B}}^{T}(t)\mathbf{S}_{\infty}&=\mathbf{Q}^{\dag}_{\infty}\mathbf{Q}_{\infty}\mathring{{\bf B}}(t)\mathbf{Q}^{-1}_{\infty}\mathbf{Q}_{\infty}+\left(\mathbf{Q}^{-1}_{\infty}\mathbf{Q}_{\infty}\right)^{\dag}\mathring{{\bf B}}^{T}(t)\mathbf{Q}^{\dag}_{\infty}\mathbf{Q}_{\infty} \\
                                                                                      &=\mathbf{Q}^{\dag}_{\infty}\left[\mathbf{Q}_{\infty}\mathring{{\bf B}}(t)\mathbf{Q}^{-1}_{\infty}+\left(\mathbf{Q}_{\infty}\mathring{{\bf B}}(t)\mathbf{Q}^{-1}_{\infty}\right)^{\dag}\right]\mathbf{Q}_{\infty} \\
                                                                                      &=\mathbf{Q}^{\dag}_{\infty}\left[\mathbf{\Lambda}_{\infty}+{\mathbf{\Lambda}}^\star_{\infty}+\mathbf{D}^{-1}\mathbf{R}_{\infty}\mathbf{D}+\left(\mathbf{D}^{-1}\mathbf{R}_{\infty}\mathbf{D}\right)^{\dag}+\mathbf{\Delta}(t)+\mathbf{\Delta}^{\dag}(t)\right]\mathbf{Q}_{\infty},
 \end{aligned}
\end{equation*} 
where
$\mathbf{\Delta}(t)=\mathbf{Q}_{\infty}(\mathring{\mathbf{B}}(t)-\mathring{\mathbf{B}}_{\infty})\mathbf{Q}^{-1}_{\infty}$ and the star denotes the complex conjugate.
Now, given that the eigenvalues $\lambda_{\infty}$ have negative real
part, 
then the constants on the
diagonal of $\mathbf{D}$ and the matrix $\mathbf{\Delta}(t)$ can be
chosen to be small enough so that there exists a
$\delta_{1}>\delta>0$, such that (see also Theorem 2.2 of \cite{KL98})
\[
 \mathbf{S}_{\infty}\mathring{{\bf B}}(t)+\mathring{{\bf B}}^{T}(t)\mathbf{S}_{\infty}\leq -2\delta_{1}\mathbf{S}_{\infty},
\]
for all $t\geq T$, which proves the lemma.
\end{proof}
Since $\mathbf{S}_{\infty}$ is positive definite, we can use this
matrix to define a new norm 
\[
\|\breve{\mathbf{u}}\|^{2}_{S_{\infty}(\mathbb{T}^{n})} \equiv \langle\breve{\mathbf{u}},\mathbf{S}_{\infty}\breve{\mathbf{u}}\rangle,
\]
which is equivalent to the usual  $L^{2}(\mathbb{T}^{n})$ norm, in the sense that 
\begin{equation}
\label{newnorm1}
\frac{1}{\tilde{C}} \|\breve{\mathbf{u}}\|^{2}_{L^{2}(\mathbb{T}^{n})}\leq \|\breve{\mathbf{u}}\|^{2}_{S_{\infty}(\mathbb{T}^{n})}\leq \tilde{C} \|\breve{\mathbf{u}}\|^{2}_{L^{2}(\mathbb{T}^{n})}\quad,\quad \tilde{C}\geq1.
\end{equation}
\begin{thm}
\label{StabilityTheorem}
 Consider the initial-value problem~\eqref{IVP_nonlinear_pert}, where
$\mathring{\mathbf{A}}^{0}$, $\breve{\mathbf{A}}^{0}$,
$\mathring{\mathbf{A}}^{j}$ and $\breve{\mathbf{A}}$ are symmetric
matrices and $\mathring{\mathbf{A}}^{0}$ is positive definite. If
Assumptions 1-3 hold, then there exists $T>0$, and $\epsilon_0>0$ such
that for $0\leq \epsilon< \epsilon_0$, and all $t\geq T$, a unique
solution of the nonlinear perturbations exists and decay exponentially
to zero, in an norm equivalent to the $H^{k}(\mathbb{T}^n)$ norm, if
$k\geq n+2$.
\end{thm}
\begin{proof}
The proof is based on Lemma~\ref{AsymptEigenValue} and
standard estimates for nonlinear symmetric hyperbolic systems
following the general strategy of Section 6.4.1 in \cite{KREISSBOOK}. Accordingly, one begins by applying the matrix $\mathbf{S}_{\infty}$ to \eqref{IVP_nonlinear_pert} and differentiating the resulting equation,
 with respect to the spatial variables, to obtain
\begin{equation}
\begin{aligned}
 \langle\partial^{\alpha}_{x}\breve{\mathbf{u}},\left[\bar{\mathring{\mathbf{A}}}^{0}+\epsilon\bar{\breve{\mathbf{A}}}^{0}\right]\partial_{t}\partial^{\alpha}_{x}\breve{\mathbf{u}}\rangle
 =& \langle\partial^{\alpha}_{x}\breve{\mathbf{u}},\left[a^{-1}(t)\bar{\mathring{\mathbf{A}}}^{j}+\epsilon\bar{\breve{\mathbf{A}}}^{j}\right]\partial_{j}\partial^{\alpha}_{x}\breve{\mathbf{u}}\rangle
 +\langle\partial^{\alpha}_{x}\breve{\mathbf{u}},\bar{\mathring{\mathbf{B}}}(t)\partial^{\alpha}_{x}\breve{\mathbf{u}}\rangle
 +\epsilon \langle\partial^{\alpha}_{x}\breve{\mathbf{u}},R^{\alpha}\rangle,
\end{aligned}
\label{Toesti}
\end{equation}
where 
\begin{equation*}
 R^{\alpha}=-\left[\partial^{\alpha}_{x}\left(\bar{\breve{\mathbf{A}}}^{0}\partial_{t}\breve{\mathbf{u}}\right)-\bar{\breve{\mathbf{A}}}^{0}\partial_{t}\partial^{\alpha}_{x}\breve{\mathbf{u}}\right]
            +\left[\partial^{\alpha}_{x}\left(\bar{\breve{\mathbf{A}}}^{j}\partial_{j}\breve{\mathbf{u}}\right)-\bar{\breve{\mathbf{A}}}^{j}\partial_{j}\partial^{\alpha}_{x}\breve{\mathbf{u}}\right]
            +\partial^{\alpha}_{x}\left(\bar{\breve{\mathbf{B}}}\breve{\mathbf{u}}\right)
\end{equation*}
denote \emph{lower order terms}, i.e. terms involving derivatives of $\breve{\mathbf{A}}^{j}$, $\breve{\mathbf{B}}$ 
and $\breve{\mathbf{u}}$, up to order $|\alpha|$, and where a bar denotes the matrix transformation
 \begin{equation*}
 \bar{\mathbf{M}} \equiv \left(\mathbf{S}_{\infty}\mathbf{M}+\mathbf{M}^{T}\mathbf{S}_{\infty}\right),
 \end{equation*}
for any real matrix $\mathbf{M}$. 

Now, note that, for sufficiently small $\epsilon$ and, in view of Assumption 1 and Assumption 2, 
$\left[\bar{\mathring{\mathbf{A}}}^{0}+\epsilon\bar{\breve{\mathbf{A}}}^{0}\right]$ 
is positive and bounded, so that \eqref{IVP_nonlinear_pert} can be used to replace 
$\partial_{t}\breve{\mathbf{u}}$ in the previous equation for $R^{\alpha}$. 
In addition, one can, as done in~\eqref{newnorm1}, define a new norm from $\left[\bar{\mathring{\mathbf{A}}}^{0}+\epsilon\bar{\breve{\mathbf{A}}}^{0}\right]$ which, 
up to a constant $\tilde{C}_{\epsilon}\geq1$, is equivalent to the $H^k(\mathbb{T}^{n})$ Sobolev norm.

Making use of the observations of the previous paragraph one can mimic
the arguments of Lemma 6.4.1 and Corollary 6.4.2 in \cite{KREISSBOOK} and find that for
fixed $k\geq n+2$ and $\epsilon>0$, there is a time $T_{\epsilon}>0$
depending on $\|\breve{\mathbf{u}}_{0}\|^{2}_{H^{k}(\mathbb{T}^{n})}$, but not on
higher derivatives of $\breve{\mathbf{u}}_{0}$, such that
\begin{equation}
 \sup_{T \leq t \leq T_{\epsilon}}\|\breve{\mathbf{u}}\|^{2}_{H^{k}(\mathbb{T}^{n})}\leq 2 \|\breve{\mathbf{u}}_{0}\|^{2}_{H^{k}(\mathbb{T}^{n})}.
\label{NONLIN_LOC}
 \end{equation}
This basic estimate ensures the local existence of the Cauchy problem
\eqref{IVP_nonlinear_pert} with data prescribed at $t_0=T$ with $T$
sufficiently large.

To show global existence one chooses  $T_{\epsilon}$ as large as
possible so that there are two possibilites:
either $T_{\epsilon}=\infty$ or $T_{\epsilon}<\infty$.  We shall now
give a small sketch of the proof that, for sufficiently small
$\epsilon$, one has $T_{\epsilon}=\infty$. The argument relies on the
estimates for local existence of solutions to quasilinear symmetric hyperbolic
systems (for full details we refer
to~\cite{KREISSBOOK,KS97,KL98,KOR98}). 

First, note that if ${\mathring{\mathbf{A}}}^{j}$ is symmetric then so is
$\bar{\breve{\mathbf{A}}}^{j}$. Integration by parts in \eqref{Toesti} then yields
\[
\langle \partial^{\alpha}_{x}\breve{\mathbf{u}},\bar{\breve{{\bf A}}}^{j}\partial_{j}\partial^{\alpha}_{x}\breve{\mathbf{u}}\rangle
=-\frac{1}{2}\langle \partial^{\alpha}_{x}\breve{\mathbf{u}},\left[\partial_{j}\bar{\breve{\mathbf{A}}}^{j}+(\partial_{u}\bar{\breve{\mathbf{A}}}^{j})(\partial_{j}\breve{\mathbf{u}})\right]\partial^{\alpha}_{x}\breve{\mathbf{u}}\rangle, 
\]
which, in view of Assumption \ref{AA1}, and using \eqref{sobo1}-\eqref{chain},
can be estimated as (see also \cite{KOR98})
\begin{equation*}
\begin{split}
\sum^{k}_{\alpha=0} |\langle \partial^{\alpha}_{x}\breve{\mathbf{u}},\bar{\breve{{\bf A}}}^{j}\partial_{j}\partial^{\alpha}_x\breve{\mathbf{u}}\rangle|
&\leq \tilde{C}^2 K(c,1) \sum^{k}_{\alpha=0} \left(1+\sum^{n=3}_{j=1}\|\partial_{j}\breve{\mathbf{u}}\|_{L^{\infty}(\mathbb{T}^{n})}\right)\|\partial^{\alpha}_{x}\breve{\mathbf{u}}\|^{2}_{L^2(\mathbb{T}^{n})} \\
&\leq \tilde{C}^2 M_1 \|\breve{\mathbf{u}}\|^2_{H^{k}(\mathbb{T}^{n})}\qquad\text{if}\quad k\geq1+[n/2]+|j|,
\end{split}
\end{equation*} 
with $M_1=
K(c,1)(1+C_{k,n}\|\breve{\mathbf{u}}\|_{H^{k}(\mathbb{T}^{n})})$. More
generally, let $M_{\alpha}$ denote polynomials in
$\|\breve{\mathbf{u}}\|_{H^{k}(\mathbb{T}^{n})}$ of degree $|\alpha|$
depending only on the constants $K(c,k)$ of Assumption
\ref{AA1}. Using the Sobolev inequalities and estimates based on the
chain rule, it can be shown that the lower order terms satisfy
\cite{KOR98}
\[
|\langle \partial_x^{\alpha}\breve{\mathbf{u}}, R^{\alpha}\rangle| \leq M_{\alpha}\|\breve{\mathbf{u}}\|^{2}_{H^{k}(\mathbb{T}^{n})}.
\]
Making use of these estimates in equation \eqref{Toesti}, one then obtains 
\[
\begin{aligned}
 \frac{d}{dt}\|\breve{\mathbf{u}}\|^{2}_{H^{k}(\mathbb{T}^{n})}\leq-\left(2\delta_{1}-\epsilon M\right)\tilde{C}^{2}_{\epsilon}\|\breve{\mathbf{u}}\|^{2}_{H^{k}(\mathbb{T}^{n})}\qquad\text{for}\quad k\geq n+2,
\end{aligned}
\]
where $M=\max_{\alpha}M_{\alpha}$. Exploiting the fact that the estimate \eqref{NONLIN_LOC} holds for $t\in[T,T_{\epsilon})$, and using
it to estimate $M$, it follows that we can choose $\epsilon$
sufficiently small such that \cite{KOR98}
\begin{equation*}
 \sup_{T \leq t \leq T_{\epsilon}}\|\breve{\mathbf{u}}\|^{2}_{H^{k}(\mathbb{T}^{n})}\leq \|\breve{\mathbf{u}}_{0}\|^{2}_{H^{k}(\mathbb{T}^{n})},
\end{equation*}
which improves \eqref{NONLIN_LOC}.
Whence, a simple continuation and contradiction argument gives the
desired global existence with exponential decay rate. 
\end{proof}
In the next section, we will see how this result can be used to show
the asymptotic exponential decay in time of nonlinear perturbations of
FLRW spacetimes containing a nonlinear scalar field.
\subsubsection{Main theorem}
In this section, we show how Assumptions \ref{ASS1}-\ref{ASS3} are satisfied for our particular problem, 
so that Theorem~\ref{StabilityTheorem} applies.

Assumption \ref{ASS1} follows by direct inspection of the perturbation matrices $\breve{\mathbf{A}}^{\mu}$ and 
$\breve{\mathbf{B}}$. Another way to see this is to notice that since $\breve{\mathbf{A}}^{\mu}$ and 
$\breve{\mathbf{B}}$ arise from the evolution 
system \eqref{EFsf} through the linearisation procedure, and since the
matrix-valued functions $\mathbf{A}^\mu$ and the vector-valued function $\mathbf{B}$ of \eqref{EFsf} 
have smooth dependence on the solution, it follows that the matrices $\breve{\mathbf{A}}^{\mu}$ and $\breve{\mathbf{B}}$ 
also have smooth dependence of the background and perturbation variables, and thus,  
Assumption~\ref{ASS1} is satisfied. 

Since $\mathring{\mathbf{A}}^{j}$ and $\mathring{\mathbf{B}}$ have a smooth dependence on $\mathring{{\bf u}}$, it follows that 
Assumption \ref{ASS2} is satisfied if the background solution $\mathring{{\bf u}}(t)$ 
is bounded  and converges to a finite limit $\mathring{{\bf u}}_{\infty}$, as $t\rightarrow+\infty$. 
It is then crucial to first understand the global behavior of the background solutions.  
The system \eqref{evol_background}-\eqref{Friedmannequation} has been studied by Rendall in a series of works 
\cite{Ren04,Ren05,Ren07}, see also \cite{Bie06}.  
For completeness, we state an important theorem which will be used in the sequel and refer to~\cite{Ren04} for
details.
\begin{Assumption}
\label{ASS4}
The scalar field potential $\mathcal{V}(\mathring{\phi})$ satisfies the  conditions:
\begin{enumerate}
\item  $\mathcal{V}(\mathring{\phi})\ge \mathring{\mathcal{V}}_0>0$, with  $\mathring{\mathcal{V}}_0$ a constant.
\item  $\mathcal{V}'(\mathring{\phi})$ is bounded on any interval on which $\mathcal{V}(\mathring{\phi})$ is bounded;
\item  $\mathcal{V}'(\mathring{\phi})$ tends to a limit, finite or infinite as $\mathring \phi$ tends to $- \infty$ or $+ \infty$.
\end{enumerate}
\end{Assumption}
Making use of this assumption one has the following:
\begin{thm} [Rendall]
\label{thm-rendall-1}
Consider a smooth spatially flat homogeneous and isotropic solution of the Einstein equations with a nonlinear scalar field 
with a positive potential $\mathcal{V}(\mathring{\phi})$ satisfying Assumption \ref{ASS4}. If the solution is initially expanding and exists 
globally to the future, then as $t\rightarrow+\infty$, it follows that $\mathring{\psi}\rightarrow0$, and 
$\mathcal{V}(\mathring{\phi})$ converges to 
some positive constant $\mathring{\mathcal{V}}_\infty$. Moreover $\mathcal{V}^{\prime}(\mathring{\phi})\rightarrow0$, and 
\[
\mathring{H}\rightarrow\sqrt{\frac{\mathring{\mathcal{V}}_\infty}{3}}.
\]
\end{thm}
As remarked in~\cite{Ren04}, conditions 2 and 3 of Assumption
\ref{ASS4} are satisfied by a general class of potentials, while
condition 1 is more restrictive, as it imposes that the potential has a strictly
positive lower bound. If a background solution is as in Theorem~\ref{thm-rendall-1}, 
then as $t\rightarrow+\infty$, $\mathring{\phi}$ converges to a (isolated) critical point of the potential  
(possibly at infinity), with $\mathcal{\mathring V}_{\infty}>0$, which is interpreted as an
effective positive cosmological constant. In that case, the deceleration
parameter $q=-1-\frac{d\mathring{H}}{dt}/\mathring{H}^{2}$ tends to $-1$ and the scale factor
grows at an exponential (accelerated) rate. In turn, the metric
locally asymptotically approaches the de Sitter metric. Thus, Theorem~\ref{thm-rendall-1}
constitutes a generalisation of the well-known theorem by
Wald~\cite{Wald, Lee}.

In subsequent work~\cite{Ren05}, Rendall considered positive potentials for which 
$\mathcal{V}(\mathring{\phi})\rightarrow0$ as $\mathring{\phi}\rightarrow\infty$. 
In that case, $\mathring{H}\rightarrow0$ and the rate of decay and
convergence of the above quantities is no longer exponential (in
sychronous time $t$). These results, show that there is a fairly
general class of potentials for which Assumption~\ref{ASS2} is
satisfied. There are, however, classes of potentials for which this is
not the case, as shown in~\cite{Ren07}.

Let us now consider the characteristic polynomial of the matrix $\mathring{\mathbf{B}}(t)$ in the system \eqref{linearised-system},
\begin{equation}
\begin{aligned}
& \left[\lambda^{3}+9\mathring{H}\lambda^{2}+\left(\mathring{\mathcal{V}}^{\prime\prime}+18\mathring{H}^{2}-6\mathring{\psi}^{2}\right)\lambda+\left(6\mathring{H}\mathring{\mathcal{V}}^{\prime\prime}+3\mathring{\psi}^{2}\left(\frac{\mathring{\mathcal{V}}^{\prime}}{\mathring{\psi}}\right)\right)\right]\times\left[\lambda^{2}+6\mathring{H}\lambda+\left(8\mathring{H}^{2}-2\mathring{\psi}^{2}\right)\right]^{3}\times \\
&\times\left[\lambda^{2}+\frac{38}{3}\mathring{H}\lambda+8\mathring{H}^{2}-2\mathring{\psi}^{2}\right]\times\left[\lambda^{2}+38\mathring{H}\lambda+72\mathring{H}^{2}-18\mathring{\psi}^{2}\right]\times \\
&\times\left[\lambda^{2}-\left(\mathring{H}+2\left(\frac{\mathring{\mathcal{V}}^{\prime}}{\mathring{\psi}}\right)\right)\lambda-\left(2\mathring{H}^{2}+2\mathring{H}\left(\frac{\mathring{\mathcal{V}}^{\prime}}{\mathring{\psi}}\right)+\frac{d\mathring{H}}{dt}\right)\right]^{3}\times 
\left[\lambda+2\mathring{H}\right]^{4}\times\left[\lambda+\frac{2}{3}\mathring{H}\right]^{3}\times\left[\lambda+\mathring{H}\right]^{18}.\nonumber
\end{aligned}
\end{equation}
In order to obtain conditions from the characteristic polynomial, we will make use of the {\it Li\'enard-Chipart theorem}. 
The latter gives necessary and sufficient conditions for a polynomial with real coefficients to have roots
with negative real part, see e.g. \cite{RS02}. The conditions for the
negativity of the real part of the  eigenvalues are found to be
\begin{equation}
\begin{aligned}
\label{cond1}
                                     \mathring H&>0, \\
\mathcal{\mathring V}-\mathring H^{2}=2\mathring H^{2}+\frac{d\mathring H}{dt}&>0, \\
  -\left(2\frac{\mathcal{\mathring V}^{\prime}}{\mathring{\psi}}+\mathring H\right)&>0, \\
  -\left(2\mathring H\frac{\mathcal{\mathring V}^{\prime}}{\mathring{\psi}}+2\mathring H^{2}+\frac{d\mathring H}{dt}\right)=-\left(2\mathring{H}\frac{\mathcal{\mathring V}^{\prime}}{\mathring{\psi}}+\mathcal{\mathring V}-\mathring{H}^{2}\right)&>0, \\
  \mathcal{\mathring V}^{\prime\prime}+12\left(\mathcal{\mathring V}-\frac{3}{2}\mathring H^{2}\right)&>0, \\
  6\mathring{H}\mathcal{\mathring V}^{\prime\prime}+3\mathring \psi^{2}\frac{\mathcal{\mathring V}^{\prime}}{\mathring \psi}&>0, \\
  3\mathring{H}\left[\mathcal{\mathring V}^{\prime\prime}+36\left(\mathcal{\mathring V}-\frac{3}{2}\mathring H^{2}\right)\right]-3\mathring \psi^{2}\left(\frac{\mathring{\mathcal{V}}^{\prime}}{\mathring \psi}\right)&>0.
\end{aligned}
\end{equation}
The first condition implies that the background solution is ever expanding and, in particular, since $\mathring{H}$ is monotonically decreasing  
it must converge to a strictly positive value $\mathring{H}_{\infty}>0$, as $t\rightarrow+\infty$. 
Let us, therefore, assume that the scalar field potential $\mathcal{V}(\mathring{\phi})$ satisfies Assumption~\ref{ASS4}, so that Theorem~\ref{thm-rendall-1} applies. Further assuming that $\mathring{\mathcal{V}}^{\prime}/\mathring{\psi}$ converges to a constant (see Remark~\ref{remark5}), it follows that 
$\mathring{\mathbf{A}}^{j}(t)\rightarrow 0$ and $\mathring{\mathbf{B}}(t)\rightarrow \mathring{\mathbf{B}}_\infty$, as $t\rightarrow+\infty$.
Moreover, conditions \eqref{cond1} at infinity reduce to
\begin{equation}
\label{cond1-at-infinity}
 \mathcal{\mathring V}_{\infty}>0,\qquad -\left(\frac{\mathcal{\mathring V}^{\prime}}{\mathring\psi}\right)_{\infty}>\sqrt{\frac{\mathcal{\mathring V}_{\infty}}{3}},\qquad\mathcal{\mathring  V}^{\prime\prime}_{\infty}>0.
\end{equation}
If these conditions are satisfied, then there exists a $\delta_\infty>0$ such that the eigenvalues $\lambda_\infty$ of 
$\mathring{\mathbf{B}}_\infty$, satisfy $\text{Re}(\lambda_\infty)\leq -\delta_\infty$.
Thus, Assumptions~\ref{ASS2} and~\ref{ASS3} are satisfied and Lemma~\ref{AsymptEigenValue} follows. 
We summarise our results in the following theorem:
\begin{thm}
\label{Theorem:Main}
Consider an initially expanding spatially flat homogeneous and isotropic solution of the Einstein-nonlinear scalar field 
system, existing globally to the future and satisfying \eqref{cond1-at-infinity}. 
Then, this solution is future asymptotically stable in the sense that there is a time $T>0$ and an $\epsilon_{0}>0$ such that, 
for  $\epsilon_{0}>\epsilon\geq0$, and all $t\geq T$, the solutions to the evolution system \eqref{IVP_nonlinear_pert} satisfy 
\begin{equation*}
\lim_{t\rightarrow\infty}\|\breve{\mathbf{u}}\|_{H^{5}(\mathbb{T}^{3})}=0
\end{equation*}
 with an exponentially decay rate.
\end{thm}
Our formulation
of the Einstein evolution equations as a first order system and our gauge
choice led to the additional condition 
\[
-(\mathcal{\mathring
  V}^{\prime}/\mathring\psi)_{\infty}>\sqrt{\mathcal{\mathring
    V}_{\infty}/{3}}
\]
 which is not present in the well know work of
Ringstr\"om \cite{Rin08}, where the harmonic gauge has been used. 
In principle, it is possible to use our approach to analyse
the FLRW open ($k=-1$) case and the ever-expanding Bianchi 
models. This can be done at the expense of further lower order
terms in \eqref{Toesti}. 
Finally, through a
suitable change of time variable, it is also possible to prove
the nonlinear stability of power-law inflation as in \cite{Rin09}.
%
\section*{Acknowledgements}
AA and FM were supported by projects PTDC/MAT/108921/2008, CERN/FP/123609/2011 and \\PTDC/MAT-ANA/1275/2014
and by CMAT, Univ. Minho, through FEDER Funds COMPETE and FCT Project Est-OE/MAT/UI0013/2014.
AA thanks the Relativity Group at the School of Mathematical Sciences,
Queen Mary, University of London, for their warm hospitality while most of
this work was done and FCT for grant SFRH/BD/48658/2008. JAVK was supported by an EPSRC Advanced Research Fellowship
and by a project research grant from the Leverhulme Trust  (F/07
476/AI). JAVK thanks the Centre of Mathematics of the University of
Minho for its hospitality.

\end{document}